\documentstyle[12pt]{article}
\begin{document}
\title{Effective Actions for Spin $0,\frac{1}{2},1$ in Curved Spacetimes}
\author{Frank Antonsen and Karsten Bormann\\ Niels Bohr Institute\\ Blegdamsvej
17\\ DK-2100 Copenhagen \O}
\date{}
\maketitle
\begin{abstract}
We calculate the effective potentials for scalar, Dirac and Yang-Mills
fields in curved backgrounds using a new method for the determination of the
heat kernel involving a partial resummation of the Schwinger-DeWitt series. 
Self-interactions are treated both to one loop order as usual
and slightly beyond one-loop order by means of a mean-field approximation. 
The new
approach gives the familiar result for scalar fields, the Coleman-Weinberg 
potential plus corrections such as the leading-log terms, but the 
actual calculation is much faster. 
We furthermore show how to go systematically to higher loop
order. The Schwarzschild spacetime is used to exemplify the procedure.\\
Next we consider phase transitions and we show that for a classical critical
point to be a critical point of the effective potential too, certain 
restrictions
must be imposed on as well its value as on the form of the classical potential
and the background geometry.
We derive this extra condition for scalar fields with arbitrary self couplings
and comment on the case of fermions and gauge bosons. Critical points of
the effective action which are not there classically are also discussed.\\
The renormalised energy-momentum tensor for a
scalar field with arbitrary self-interaction and non-minimal coupling to
the gravitational background is calculated to this improved one-loop order as 
is the resulting conformal anomaly. Conditions for the violation of energy
conditions are given.  \\
All calculations are performed in the case of 
$d=4$ dimensions.
\end{abstract}

\section{Introduction}
The Coleman-Weinberg formula for the one-loop order
effective potential of a $\phi^4$ theory has many applications, 
e.g. to inflation and to the study of the
standard model of particle physics. But for calculations in curved spacetime,
such as more realistic inflationary scenarios and the study of the early
universe, one should take curvature into account.\\
 In this paper we will use the
heat kernel method for finding the effective potential as described in e.g. 
the textbook by Ramond,
\cite{Ramond}, but this time in curved spacetime using methods developed by 
the authors \cite{Casimir}. We should emphasise that even 
the simple approach of the first part of this
paper can be used to go beyond one-loop order, as we can easily include
higher order derivatives of the curvature and of the classical fields corresponding
to quantum corrections to the kinetic part of the effective action. In fact, we 
will argue that some of the terms in our final expression corresponds to a 
summation of leading log-terms. 
Furthermore, the mean-field approach, to which we resort in order to extend 
the results to the case of self-interacting fields, is essentially 
a non-perturbative approximation to the full effective action, as it allow us to
perform the functional integral. This mean-field approach can be iterated 
to reach, in principle, any accuracy desired. We finally compare our method
to that of other authors. For generality, we let the classical potential of 
the scalar field be completely arbitrary.\\
The next step is to treat Dirac fermions, and it is shown how to relate the 
effective actions
for such spin-$\frac{1}{2}$ fields to that of the non-minimally coupled 
scalar field. Subsequently the
renormalised mass is found. Comments on Weyl fermions are also 
made, as are 
non-minimally coupled fermions, i.e., fermions coupling to the background 
torsion, and spinors coupling to Yang-Mills fields.\\
Thirdly we consider Yang-Mills fields in curved spacetime, and we present 
the effective
action for this case too, both to one loop (or higher) order and in the 
mean field approach. In both
cases can the ghost contribution be related to the effective action for a free
scalar field.\\
Next we show how, using a prescription for propagators, 
\cite{prop}, we can in principle go to any loop order. The relationship between
the Green's function and the heat kernel is used to write down an explicit
formula for the second-loop order contribution to the effective 
action in a certain approximation.\\
We also discuss phase transitions and critical points, both of the original
classical potential and the resulting effective one. In particular we see when 
a classical critical point is also a critical point of the effective 
potential. It turns out that this leads to restrictions not only on the 
couplings but also on the background geometry.\\
From the effective action we can then also find the 
renormalised energy-momentum tensor. It turns out that this
in general violates the weak energy condition, and a 
measure for this violation is found. \\
At the end, we provide a discussion and an outlook.

\subsection*{Contents}
Part I (Theory)\\
1. Introduction\\
2. Determining the curved space Coleman-Weinberg Potential\\
3. The Mean-field Approach\\
3.1. Example: Schwarzschild Spacetime.\\
4. Fermions\\
4.1. Free Dirac Fermions\\
4.2. On Weyl Fermions\\
4.3. On Coupling to Yang-Mills Fields and Torsion\\
4.4. Example: Schwarzschild Spacetime\\
5. Yang-Mills Fields\\
5.1. Mean-field Approximation and Symmetry Restoration\\
5.2. Example: Schwarzschild Spacetime\\
6. Beyond One Loop Order\\
Part II (Applications)\\
7. Quantum Modification of Classical Critical Points\\
7.1 Spinor and Vector Fields\\
8. Quantum Critical Points\\
9. The Energy-Momentum Tensor\\
9.1. Conformal Anomaly\\
9.2. Violation of Energy Conditions\\
10. Discussion and Conclusion

\section{Determining the Curved Space Coleman-Weinberg Potential}
As shown in e.g. Ramond \cite{Ramond}, the effective 
potential to one-loop order can be found rather quickly by the heat kernel 
method:
\begin{equation}
    \int V_{\rm eff}(\phi_{\rm cl})\sqrt{g}d^4x = 
    -\frac{1}{2}\zeta_{-\Box+m^2+V''(\phi_{\rm cl})}'(0)
\end{equation}
where the zeta function is calculated assuming a constant configuration and
where $\phi_{\rm cl}$ denotes the classical field with $V''= \frac{\partial^2 V}{\partial
\phi_{\rm cl}^2}$. In curved space this must be generalised by determining the 
heat kernel of the curved space scalar field operator (from which one
can calculate the zeta-function), $G$, which thus obeys
\begin{equation}
    \left[-\Box+m^2+V''(\phi_{\rm cl})+\xi R\right]
      G(x,x';\sigma)  
    =\frac{\partial }{\partial \sigma}G(x,x';\sigma)
\end{equation}
where $\Box$ denotes the curved spacetime d'Alembertian
\begin{equation}
    \Box = \frac{1}{\sqrt{g}}\partial _\mu \sqrt{g}g^{\mu\nu}\partial _\nu = 
    \Box_0+\frac{1}{e}e^m_\mu(\partial _m(ee^\mu_a))\partial ^a
\end{equation}
with
\begin{equation}
    \Box_0 = \eta^{mn}\partial _m\partial _n
\end{equation}
and where we have introduced vierbeins $e^m_\mu$ (i.e. $g_{\mu\nu}=e_\mu^m 
e_\nu^n
\eta^{mn}$ and $\eta^{mn} = g^{\mu\nu}e_\mu^me_\nu^n$, $e=\det(e^m_\mu)=
\sqrt{g}=\sqrt{\det(g_{\mu\nu})}$).\\ 
The computation of the coefficients of the heat kernel is
rather straightforward.
\footnote{One should notice that most of the calculations to follow
don't really assume $\phi_{\rm cl}=const.$, only in the final result for the
renormalised mass and coupling constant is this assumption needed. \\
One should furthermore
notice that although the result quoted here is for $x=x'$ we can actually find
the heat kernel to arbitrary order of accuracy even for $x\neq x'$. The heat
kernel would then depend on $\Delta(x,'x)$, the geodesic distance squared, 
as well as on integrals of powers of the curvature scalar and its derivatives 
along the geodesic from $x$ to $x'$, assuming such a curve exists 
\cite{prop}.} The equation to be solved is (where $f_0$ is found from (2))
\begin{equation}
	\left(\Box_0+f_0(x)\right) G(x,x';\sigma) = 
	-\frac{\partial}{\partial\sigma}G(x,x';\sigma)
\end{equation}
subject to the boundary condition
\begin{equation}
	\lim_{\sigma\rightarrow 0}G(x,x';\sigma) = \delta(x,x')
\end{equation}
We solve this equation by writing 
\begin{equation}
	G(x,x';\sigma) = G_0(x,x';\sigma) e^{-T(x,x';\sigma)}
\end{equation}
with $G_0$ the heat kernel of $\Box_0$, which in $d=4$ dimensions is
\begin{equation} 
	G_0=\frac{\Delta_{\rm vvm}}{(4\pi\sigma)^2}
	e^{-\frac{\Delta(x,x')}{4\sigma}}
\end{equation}
with $\Delta(x,x')=(\int ds)^2$ the square of the geodesic 
distance (the so-called Synge world function) and 
\begin{displaymath}
	\Delta_{\rm vvm} \equiv \frac{\det (\partial_\mu\partial_\nu\Delta)}
	{\sqrt{g(x)g(x')}}
\end{displaymath}
being the Van Vleck-Morette determinant.\footnote{Fortunately, we only 
need these
quantities evaluated at the diagonal $x=x'$ for the present purposes, which 
simplifies matters a lot.} Doing this leads to the following 
equation for $T$
\begin{equation}
	\Box_0 T + (\partial T)^2 +f_0 = -\frac{\partial}{\partial\sigma}T
\end{equation}
subject to the condition
\begin{equation}
	\lim_{\sigma\rightarrow 0} T(x,x';\sigma) = 0
\end{equation}
Taylor expanding $T$,
\begin{equation}
	T(x,x';\sigma) = \sum_{n=0}^\infty \tau_n(x,x')\sigma^n
\end{equation}
with $\tau_0=0$ because of the initial value condition, we get the following 
recursion relation when $x=x'$
\begin{equation}
	\Box_0\tau_n + \sum_{k=0}^n \partial_m\tau_k\partial^m\tau_{n-k}
	-(n+1)\tau_{n+1} \qquad n\geq 2
\end{equation}
From the heat kernel equation it furthermore follows that 
$\tau_1(x,x) = f_0(x)$. Finding the higher coefficients is then trivial,
\cite{Casimir}.\\
We must emphasise that these coefficients differ from the usual 
Schwinger-DeWitt
ones, \cite{BD}, a difference due to the different nature of the 
two expansions used.\\ 
One 
should also note that, contrary to the Schwinger-DeWitt case, the coefficients 
given by our expansion are relatively straightforward to obtain explicitly. 
The expression (7) can actually be viewed as a partial resummation of the 
Schwinger-DeWitt series. In fact, the series is the so-called
cumulant of the (divergent/asymptotic) series of Schwinger and DeWitt,
which is strictly speaking only valid for $\sigma$ small.\footnote{If $\sum
a_n s^n$ is a divergent/asymptotic series the {\em cumulant} is given by a
series $\exp(\sum b_ns^n)$.} Our series also
has better convergence properties due the pressence of an $e^{-(m^2+\xi R)
\sigma}$-term, which is in itself an indication of an implicit summation
of leading log-terms, \cite{exiR}. 
Besides the different expansions used, the major cause for simplification 
lies in 
the usage of vierbeins and hence co-moving coordinates. This being so,
it turns out that one can actually rather easily write down expressions for
the coefficients in our series, whereas one can in general only compute the
first few in the Schwinger-DeWitt case.\\
Thus, we arrive at
\begin{eqnarray}
    G(x,x;\sigma)  &=& (4\pi\sigma)^{-2}e^{-\tau_1(x) \sigma - \tau_2(x)\sigma^2
    -\tau_3(x)\sigma^3+...}\nonumber\\
    &\approx & (4\pi\sigma)^{-2}e^{-\tau_1(x)\sigma}\left(1-\tau_2(x)\sigma^2
    -\tau_3(x)\sigma^3+...\right)\nonumber\\
    &\equiv& (4\pi\sigma)^{-2} e^{
    -\tilde{f}_0\sigma}\left(1+\frac{1}{2}\Box_0f_0\sigma^2-\frac{1}{3}
      (\partial_p f_0)^2\sigma^3+...\right) \label{eq:heat0}
\end{eqnarray} 
where
\begin{eqnarray}
    f_0 &=& m^2+V''(\phi_{\rm cl})+\xi R+{\cal E}\\
    \tilde{f}_0 &=& m^2+V''(\phi_{\rm cl})+\xi R+2{\cal E}
\end{eqnarray}
with $\cal E$ containing vierbeins and their derivatives only
\begin{equation}
    {\cal E}=\frac{1}{2}\partial ^a\left(\frac{1}{e}e^m_\mu(\partial _m(ee^\mu_a))
      \right)+\frac{1}{4}\left(\frac{1}{e}e^m_\mu(\partial _m(ee^\mu_a))\right)^2
\end{equation}
In formula (\ref{eq:heat0}) higher order terms have been
omitted, i.e., the result is valid for $R$ and $\phi_{\rm cl}^2$ strong but 
sufficiently slowly
varying in order for us to discard third and higher derivatives. It should be
noted that that is the only approximation we have made. Furthermore, one could
in principle include higher derivatives -- it is not difficult to find the
corresponding coefficients -- but the integral over $\sigma$ which we will
ultimately perform becomes difficult to make analytically.\\ 
With this expression for the heat kernel it is thus possible to 
calculate the zeta function (see e.g. 
\cite{Ramond,Casimir,EffAct})
\begin{eqnarray}
    \hspace{-5mm}\zeta(s) &=& \frac{1}{\Gamma(s)}\int_0^\infty d\sigma 
      \sigma^{s-1}\int G(x,x;\sigma) d^4x\\
    &=& (4\pi)^{-2}\left(\tilde{f}_0^{2-s}\frac{\Gamma(s-2)}{\Gamma(s)}
    +\frac{1}{2}\Box_0f_0\tilde{f}_0^{-s}-\frac{1}{3}(\partial _p f_0)^2
    \tilde{f}_0^{-1-s}\frac{\Gamma(s+1)}{\Gamma(s)}+...\right)\nonumber\\
\end{eqnarray}
and we finally arrive (from equations (11,14,15,18)) at 
\begin{eqnarray}
    V_{\rm eff} &=& \frac{1}{2}m^2\phi_{\rm cl}^2+V(\phi_{\rm
    cl})+\frac{1}{64\pi^2}(m^2+V''(\phi_{\rm cl})+\xi R+2{\cal
    E})^2\times\nonumber\\
    &&\left[\ln\left(m^2+V''(\phi_{\rm cl})+\xi R+2{\cal E})
    \right)-\frac{3}{2}\right]+
           \frac{1}{64\pi^2}\Box_0\left(V''(\phi_{\rm cl})
    +\xi R+{\cal E}\right)-\nonumber\\
    &&\frac{1}{96\pi^2}(m^2+V''(\phi_{\rm cl})+\xi R+{\cal E})^{-1}
    \left(\partial _p\left(V''(\phi_{\rm cl})
    +\xi R+{\cal E}\right)\right)^2+\frac{1}{2}\xi R\phi_{\rm cl}^2\nonumber\\
	\label{eq:veff}
\end{eqnarray}
The quantity $\cal E$ appears because we are using vierbeins and 
hence do not
have a coordinate basis, $[\partial_m,\partial_n]\neq 0$, and 
because $\Box_0$ is
not the true d'Alembertian.\\
Notice that this result, (\ref{eq:veff}), holds even for $\phi_{\rm cl}$ not 
constant. This means
that even though (1) is only valid to one-loop order we can get some of the
quantum corrections to the kinetic part, which otherwise belong to higher loop
order. Of course, we cannot expect to get all the corrections this way, but
we may get some indication of the nature of the leading terms.\\
From (\ref{eq:veff}) we can obtain the renormalised values of the mass and of 
the coupling
constant, putting $\phi_{\rm cl}=constant$ and assuming, for sake of
argument, 
$V(\phi_{\rm cl}) = \frac{\lambda}{4!}\phi_{\rm cl}^4$:
\begin{eqnarray}    
    \xi_{\rm ren} &\equiv & \left.\frac{d^3 V_{\rm eff}}{d\phi_{\rm cl}^2 
	dR}\right|_{\phi_{\rm cl}=0}\nonumber\\
	&=& \xi +\frac{\lambda}{(32\pi^2}\xi\left[\ln(m^2+\xi R+2{\cal E})
	-1\right]+\frac{\lambda\xi^2}{32\pi^2}\frac{1}
	{m^2+\xi R+{\cal E}}-\nonumber\\
	&&\qquad \frac{\lambda\xi^2}{48\pi^2}
	(m^2+\xi R+2{\cal E})^{-3}(\partial_p(\xi R+{\cal E}))^2+\nonumber\\
	&&\qquad\qquad \frac{\lambda\xi}{48\pi^2}(m^2+\xi R+2{\cal E})^{-2}
	\partial^p{\cal E}\partial_p(\xi R+{\cal E})\\
 m^2_{\rm ren}&\equiv&\left.\frac{d^2V_{\rm eff}}{d\phi_{\rm cl}^2}
    \right|_{\phi_{\rm cl}=0}-\xi_{\rm ren}R\nonumber\\
    &=& m^2+\frac{\lambda}{32\pi^2}(m^2+2{\cal E})\left[\ln(m^2+
    \xi R+2{\cal E})-1\right]+\nonumber\\
    &&\hspace{20mm}\frac{\lambda}{96\pi^2}(m^2+\xi R+2{\cal E})^{-2}
      (\partial_p(\xi R+{\cal E}))^2-\frac{\lambda\xi^2}{32\pi^2}\frac{1}
	{m^2+\xi R+{\cal E}}+\nonumber\\
&&\hspace{20mm}\frac{\lambda\xi^2}{48\pi^2}
	(m^2+\xi R+2{\cal E})^{-3}(\partial_p(\xi R+{\cal E}))^2+\nonumber\\
	&&\qquad\qquad \frac{\lambda\xi}{48\pi^2}(m^2+\xi R+2{\cal E})^{-2}
	\partial^p{\cal E}\partial_p(\xi R+{\cal E})\\ 
    \lambda_{\rm ren} &\equiv&\left.\frac{d^4V_{\rm eff}}{d\phi_{\rm cl}^4}
    \right|_{\phi_{\rm cl}=0}\nonumber\\
    &=& \lambda+\frac{3\lambda^2}{32\pi^2}\ln(m^2+\xi R+2{\cal E})
    +\nonumber\\
    &&\hspace{30mm}
    \frac{2\lambda^2}{48\pi^2}(m^2+\xi R+2{\cal E})^{-3}(\partial _p(\xi R 
    +{\cal E}))^2
\end{eqnarray}
Similarly, by writing $\Gamma=\int d^4x\sqrt{g}(\Lambda_{\rm ren}+\kappa_{\rm ren}
R+O(R^2))$ and considering this an action for the gravitational degrees of
freedom, we get the following values for the renormalised cosmological and
Newtonian constants respectively
\begin{eqnarray}
	\Lambda_{\rm ren} &=& \frac{1}{2}m^2\phi_{\rm cl}^2+V(\phi_{\rm cl})+
	\frac{1}{64\pi^2}\Box_0(V''+\xi R+{\cal E})-\nonumber\\
	&&\qquad\frac{1}{96\pi^2}(m^2+V''+\xi R +2{\cal E})^{-1}(\partial_p(
	V''+{\cal E}+\xi R))^2+\nonumber\\
	&&\qquad \frac{1}{64\pi^2}(m^2+V''+2{\cal E})\left[\ln(m^2+V''+
	2{\cal E})-\frac{3}{2}\right]\\
	\kappa_{\rm ren} &=& \frac{1}{2}\xi\phi_{\rm cl}^2+\frac{\xi}{96\pi^2}
	(m^2+V''+2{\cal E})^{-2}(\partial_p(V''+{\cal E}+\xi R))^2+\nonumber\\
	&&\qquad \frac{\xi}{64\pi^2}\left[\ln(m^2+V''+2{\cal E})-\frac{1}{2}
	\right]
\end{eqnarray}
this shows how the masses of the fundamental fields modify the Newtonian
constant, whether this effect is testable or not is difficult to say at 
present. The quantum modifications of the coupling constants are due to (1)
the mass of the scalar field, (2) the value of $\phi_{\rm cl}$ and $R$, and (3)
the variations thereoff. Consequently, also massless scalar fields will lead
to $\lambda_{\rm ren}\neq 0$. Such effects may have implications similar to
dark matter.\\
Let us at this stage pause and compare with the results found by other 
authors. 
Hu and O'Connor, \cite{phi4}, have calculated the effective potential for a 
$\phi^4$ theory in a static, homogeneous spacetime in which the interaction 
term $\frac{1}{2}\lambda\phi^2$ can be assumed constant. They then use 
dimensional regularisation and the Schwinger-DeWitt expansion to find the 
following expressions for the change in mass and
couplings constants to one-loop order:
\begin{eqnarray*}
	\delta m^2 &=& \frac{\hbar}{32\pi^2}\lambda m^2(\ln\frac{m^2}{\mu^2}-
	1)\\
	\delta\lambda &=& \frac{3\hbar}{32\pi^2}\lambda^2\left( 
	\ln\frac{m^2+\frac{1}{2}
	\lambda\phi^2}{\mu^2}+\frac{8}{3}-\frac{8m^2}{3}
	\frac{m^2+\frac{1}{4}\lambda\phi^2}
	{(m^2+\frac{1}{2}\lambda\phi^2)^2}\right)\\
	\delta\xi &=& \frac{\hbar}{32\pi^2}\lambda \xi \ln \frac{m^2-\xi R}{\mu^2}
\end{eqnarray*}
Comparing this with our result (we always take $\mu=1$) we notice that the 
two set of expressions agree provided we take $\phi=\phi_{\rm cl}=const.$ in 
Hu and O'Connor's result, remove the special curvature terms coming from the
noncoordinate basis (i.e., the $\cal E$-terms), assuming $R\approx const.$ and
ignoring all higher order terms. The $\phi$-term is easy to understand as 
it is a constant in their approach,
similarly one would expect $R$ to be absent in a Riemann normal coordinate 
patch to
the lowest order. As mentioned earlier $\cal E$ is a result of not using a 
coordinate frame in our case, $[\partial_m,\partial_n]\neq 0$.\\
There is something odd about the result of Hu and O'Connor, though, 
the non-minimal coupling, $\xi R\phi^2$, clearly acts like a mass term in the 
Lagrangian and one would thus expect the
renormalised mass to depend on $\xi$. In fact, even if the field was massless 
classically we would expect a curvature induced mass to turn up, but their 
result is that a massless field does not acquire any mass to one-loop order. 
This is in stark contrast to what one would expect in view of Shore's work, 
\cite{Shore}, where precisely such a curvature induced mass is responsible for 
symmetry restoration in scalar QED. Shore's result, which we
will also comment on later, is based on an Euclidean version of de 
Sitter space, namely $S^n$, and he then explicitly calculates the 
zeta function for scalar QED. In fact, he shows that a minimal coupling 
would lead to an anomalous mass term. 
The discrepancy between the work of Hu and O'Connor on the one hand and Shore 
and this paper on the other we think
rests on the approximations made by the authors of \cite{phi4}: a static, 
homogeneous spacetime using the Schwinger-DeWitt expansion in a Riemann normal 
coordinate patch. As 
mentioned earlier, the method put forward here is a partial resummation of the 
Schwinger-DeWitt series and will thus inevitably give different results. 
What all these methods agree on, of course, is the result in the flat space 
limit, where 
we must recover the Coleman-Weinberg potential at least as the leading 
contribution. Only keeping
the very lowest order terms in (\ref{eq:veff}) we recover (except for the 
quantity $\cal E$), the result by Hu and O'Connor.\\
The effective potential for a $\phi^4$-theory in some model spacetimes 
have been calculated by
a number of authors. For instance, Hu and O'Connor has also found the effective
potential in a mix-master universe.
Futamase, \cite{Futamase}, has used the $\zeta$-function technique in the 
Bianchi I spacetime for a $\phi^4$-theory with and without a coupling to an el
ectromagnetic field. The formula he finds is valid for
$\beta$ (the inhomogeneous parameter, showing the discrepency between the
homogenous Friedmann-Robertson-Walker spacetime and the inhomogeneous Bianchi 
I) small. Similarly, Berkin \cite{Berkin}, has found a formula by Taylor 
expanding the heat kernel in powers of $\beta$. His result too, is valid 
only for small $\beta$. 
These two authors agree with each other. Contrary to this, Huang \cite{Huang}, 
has used an adiabatic approximation 
and get a different result. He finds, for instance, that symmetry 
restoration (a point we will be returning to later) is possible even for
$\beta$ large, whereas the other authors only found this to be possible for 
$\beta$ small. We will not comment further on these results.\\
Ishikawa, \cite{Ishikawa}, has calculated the effective potential for a 
massless
$\phi^4$, scalar QED 
theory in a more general background, but using Riemann normal coordinates and 
only going to $O(R)$
assuming throughout that $|R/\lambda\phi^2| \ll 1$. He uses a 
renormalisation-group improved technique and gets
\begin{eqnarray*}
	V_{\rm eff, RG} &=& \frac{\lambda}{4!}\phi^4-\frac{1}{2}\xi R\phi^2 +
	(8\pi^2)^{-1}
	\frac{\lambda}{2}\phi^2\left(\ln\frac{\phi^2}{\mu^2}-\frac{25}{6}\right)+\\
	&&(8\pi)^{-2}(\frac{1}{6}-\xi)R\lambda\phi^2\left(\ln\frac{\phi^2}{\mu^2}-
	3\right)
\end{eqnarray*}
This differs from our result due to the approximations made (linearity in $R$,
no derivatives of the curvature, no $\cal E$-term).\\
In general then, the previous research have mostly been limited to the 
case $|R|\ll |\lambda\phi^2|$, in which case one gets results linear in $R$ or 
the (slightly improved) Coleman-Weinberg
potential from flat spacetime with $m^2\rightarrow m^2+\frac{1}{2}\lambda\phi^2$.
Contrary to this, we do not require $|R|$ small. For purely practical purposes 
we have, though, limited ourselves to the regime in which $\partial R,
\partial^2 R$ are small 
(compared to $R$). However, as pointed out in \cite{Casimir}, we could in 
principle remove this limitation.

\section{The Mean-Field Approach}
Only Gaussian functional integrals can be calculated in any reliable 
manner. We must consequently find a way of transforming the original 
functional integral 
defining the partition function, $Z$, into a Gaussian one. 
An improvement over the previous method can be 
obtained by writing\footnote{i.e., instead of making use of the prescription 
$\phi \rightarrow \phi_{\rm fluct}+\phi_{\rm cl}$ (perhaps keeping 
$\phi_{\rm cl}$ constant, and in any case only keeping terms 
$O(\phi_{\rm fluct}^2)$ 
when functionally integrating out the fluctuating part $\phi_{\rm fluct}$) 
for part of the interaction term we make the prescription
$\phi^4 \rightarrow \langle\phi^2\rangle \phi^2$ where 
$\langle\phi^2\rangle$ is the actual mean
value of the the field due to the propagation of virtual particles in 
curved space.}.
\begin{equation}
    {\cal L} = \frac{1}{2}\phi(-\Box+m^2+\xi R+\frac{1}{2}\lambda \langle \phi^2
    \rangle)\phi \label{eq:Lphi}
\end{equation}
where $\langle \phi^2\rangle$ is a mean field. A similar approach for 
Yang-Mills fields have been
developed in \cite{Casimir}, and we will briefly outline it for the simpler 
case 
of a scalar field. Supposing that we have
calculated $\langle \phi^2\rangle$, the heat kernel is (cf. equation (13))
\begin{equation}
    G(x,x;\sigma) = (4\pi\sigma)^{-2}e^{-\tau_1\sigma}\left(1-\tau_2\sigma^2
    -\tau_3\sigma^3-...\right) \label{eq:heatexp}
\end{equation}
with
\begin{eqnarray*}
    \tau_1 &=& m^2+\xi R +\frac{1}{2}\lambda \langle\phi^2\rangle \\
    \tau_2 &=& -\frac{1}{2}\Box_0\tau_1\\
    \tau_3 &=& \frac{1}{3}(\partial \tau_1)^2+\frac{1}{6}\Box_0^2\tau_1
\end{eqnarray*}
With a recursion relation identical in form to the one given in (12).
In the previous calculation we have ignored the $\Box_0^2\tau_1$ term in 
$\tau_3$
as this is a fourth order derivative of the curvature essentially.\\
From this one easily obtains the effective action.\\
Also, the mean field $\langle \phi^2\rangle$ is obtained to the lowest order
from
\begin{equation}
    \langle\phi(x)\phi(y)\rangle = \left(\frac{\delta^2S_0}{\delta \phi(x)\delta\phi(y)}
    \right)^{-1} = G(x,y)\label{eq:phi2}
\end{equation}
where $S_0$ is the action of $\phi$ with $\lambda=0$, i.e., only the 
kinetic and gravitational
terms are included; the self-interaction is ignored. To the next order one 
could define an
$S_1$ with $\lambda\neq 0$ and $\langle \phi^2\rangle$ given by 
(\ref{eq:phi2}), this can then be
iterated to any order of accuracy wanted -- the functional integrals are 
easily solved, as one
just needs to calculate the new coefficients $\tau_n$ in the expansion above 
of the 
heat kernel. For completeness, the first iterated coefficients are listed at 
the end of this section.\\
Now, the inverse of the second derivative of the action is the Green's function
$G(x,x')$. This too can be found from the heat kernel (see \cite{prop}) as
\begin{equation}
    G(x,x') = -\int_0^\infty d\sigma G(x,x';\sigma)
\end{equation}
which follows directly from the spectral decompositions of the heat kernel and 
the Green's function respectively.\\
Inserting the above expansion (\ref{eq:heatexp}) with $\lambda=0$ we get a 
divergent result
for $G(x,x) = \langle \phi(x)^2\rangle$ as one would expect (it is proven 
in \cite{prop} that the propagator one gets from our expression for the 
heat kernel satisfies the 
Hadamard condition and thus has the correct singularity structure as 
$x'\rightarrow x$), but the infinities can be
removed quite simply by using principal values instead (the singularity is of 
the form $\Gamma(-1)$) \cite{princip}. The final result thus becomes
\begin{equation}
    \langle \phi(x)^2\rangle_{\rm reg}^{(0)} = (4\pi)^{-2}\left[-(\gamma-1)
	\bar{\tau}_1^{(0)} +
    \bar{\tau}_2^{(0)}(\bar{\tau}_1^{(0)})^{-1}+\bar{\tau}_3^{(0)}
	(\bar{\tau}_1^{(0)})^{-2}+...\right] 
		\label{eq:meanfield}
\end{equation}
with $\bar{\tau}_n^{(0)} \equiv \lim_{\lambda\rightarrow 0}\tau_n$, and where 
we 
did not iterate the mean field.\footnote{Actually, one can think of this 
calculation as determining the contribution to the 
mean field due to the propagation
of virtual particles in curved space whereas the next iteration includes 
the self-interaction and thus, essentially, is a one or higher loop-order 
calculation (in the matter fields, this time). The superscript $(0)$ refers
to the lack of iteration.} Explicitly:
\begin{eqnarray*}
    \bar{\tau}_1^{(0)} &=& -m^2-\xi R\\
    \bar{\tau}_2^{(0)} &=& -\frac{1}{2}\xi\Box_0 R\\
    \bar{\tau}_3^{(0)} &=& -\frac{1}{3}\xi^2(\partial R)^2-\frac{1}{6}\xi 
	\Box_0^2 R\\
    &\approx & -\frac{1}{3}\xi^2(\partial R)^2
\end{eqnarray*}
and so on. Discarding third and higher derivatives of $R$ (i.e., going only 
to one 
loop order in gravity) the remaining coefficients all vanish. Thus
\begin{displaymath}
	\langle\phi^2\rangle^{(0)}_{\rm ren} = \frac{\gamma-1}{(4\pi)^2}
(-m^2+\xi R+\mbox{derivatives of $R$})
\end{displaymath}
is the formula we find for the curvature induced mean field.\\
By the definition of the Lagrangian, (\ref{eq:Lphi}), the mean field can 
be seen as a 
redefinition of the mass and the non-minimal coupling
\begin{eqnarray}
	\delta m^2 &=& \frac{\lambda}{4!(4\pi)^2}\left\{(\gamma-1)m^2+ \frac{1}{2}\xi
	\frac{\Box_0 R}{m^2+\xi R}-\frac{1}{3}\xi^2 \frac{(\partial R)^2}
	{(m^2+\xi R)^2}\right\}\\
	\delta\xi &=& \frac{\lambda}{4!(4\pi)^2}(\gamma-1)\xi
\end{eqnarray}
which shows the non-perturbative nature of this approximation very clearly (it
is non-polynomial in the coupling to the curvature, $\xi$).\\
The coefficients $\tau_n$ which enter the full heat kernel are then
\begin{eqnarray*}
    \tau_1 &=& m^2+\xi R + \frac{1}{2}\frac{\lambda}{(4\pi)^2}\left[(\gamma-1)(m^2+\xi R)
    +\frac{1}{2}\xi \Box_0 R \cdot (m^2+\xi R)^{-1}+\right.\\
    &&\qquad\left.+\frac{1}{3}\xi^2(\partial R)^2(m^2+\xi R)^{-2}\right]\\
    \tau_2 &\approx & \xi\Box_0 R+\frac{1}{2}\frac{\lambda}{(4\pi)^2}\left[
    \xi(\gamma-1)\Box_0 R+2\xi^3\Box_0 R(\partial R)^2(m^2+\xi R)^{-3}\right.\\
    &&\qquad+\left.\frac{1}{2}\xi^3(\Box_0 R)^2(m^2+\xi R)^{-2} +2\xi^4(\partial R)^4
    (m^2+\xi R)^{-4}\right]\\
    \tau_3 &\approx & \frac{1}{3}\left\{\xi \partial_pR+\frac{1}{2}
	\frac{\lambda}{(4\pi)^2}\left[
    (\gamma-1)\xi\partial_p R-\frac{1}{2}\xi\Box_0 R\partial_p R(m^2+\xi R)^{-2}
	\right.\right.\\
    &&\qquad\left.\left.-\frac{2}{3}\xi^2(\partial R)^2\partial_p R(m^2+\xi R)^{-3}
	\right]\right\}^2
\end{eqnarray*}
As this result is based on a mean field approximation it is non-perturbative, 
and iterating the process as described above, we could further increase the 
accuracy of
the calculation, it is clear, however, that this would be somewhat cumbersome,
though not at all difficult in principle. To give an example of the 
procedure, we just list the first iterated mean-field
\begin{equation}
    \langle \phi^2\rangle_{\rm reg}^{(1)} = (4\pi)^{-2}\left[-(\gamma-1)
	\bar{\tau}_1^{(1)} 
    +\bar{\tau}_2^{(1)}(\bar{\tau}_1^{(1)})^{-1}+\bar{\tau}_3^{(1)}
	(\bar{\tau}_1^{(1)})^{-2}+...\right]
	\label{eq:mean2}
\end{equation}
with
\begin{eqnarray*}
    \bar{\tau}_1^{(1)} &=& \bar{\tau}_1^{(0)}-\frac{1}{2}
	\lambda\langle\phi^2\rangle_{\rm reg}^{(0)}\\
    \bar{\tau}_2^{(1)} &=& -\frac{1}{2}\Box_0(\xi R+\lambda\langle\phi^2
	\rangle_{\rm reg}^{(0)})\\
    \bar{\tau}_3^{(1)} &\approx & -\frac{1}{3}\left(\partial(\xi R+
	\lambda\langle\phi^2\rangle_{\rm reg}^{(0)})\right)^2
\end{eqnarray*}
The procedure is now transparent. One should note, as follows from 
(\ref{eq:meanfield}), that our normalisation is such that the
mean field vanishes whenever $m=R=0$.\\
Using (\ref{eq:mean2}) in the the general expression for $V_{\rm eff}$, eq 
(\ref{eq:veff}), i.e. putting $V''(\phi_{\rm cl}) = \frac{1}{2}\lambda
\langle\phi^2\rangle^{(1)}_{\rm reg}$
we can find the new corrections to the mass and non-minimal coupling
\begin{eqnarray}
	\delta m^2 &=& \frac{\lambda}{4!(4\pi)^2}\left[(\gamma-1)(1+\lambda(\gamma-1))m^2
	+\frac{1}{2}\lambda\xi(\gamma-1)\frac{\Box_0 R}{m^2+\xi R} -
	\frac{1}{3}\lambda \xi^2(\gamma-1)\frac{(\partial R)^2}{(m^2+\xi R)^2}+\right.
	\nonumber\\
	&&\left(\frac{1}{2}\xi (1+\lambda(\gamma-1))\Box_0 R+\frac{1}{2}\lambda\xi \Box_0
	\frac{\Box_0 R}{m^2+\xi R}-\frac{1}{3}\lambda\xi^2\Box_0\frac{(\partial R)^2}
	{(m^2+\xi R)^2}\right)\times\nonumber\\
	&&\left((m^2+\xi R)(1+\lambda(\gamma-1))+\frac{1}{2}\lambda\xi\frac{\Box_0 R}
	{m^2+\xi R}-\frac{1}{3}\lambda\xi^2\frac{(\partial R)^2}{(m^2+\xi R)^2}\right)
	^{-1}-\nonumber\\
	&&\frac{1}{3}\left(\partial\left(\xi R (1+\lambda(\gamma-1))+\frac{1}{2}\lambda
	\xi\frac{\Box_0 R}{m^2+\xi R}-\frac{1}{3}\lambda\xi^2\frac{(\partial R)^2}
	{(m^2+\xi R)^2}\right)\right)^2\times\nonumber\\
	&&\left.\left((m^2+\xi R)(1+\lambda(\gamma-1))+\frac{1}{2}\lambda\xi\frac{\Box_0 R}
	{m^2+\xi R}-\frac{1}{3}\lambda\xi^2\frac{(\partial R)^2}{(m^2+\xi R)^2}\right)
	^{-2}\right] \label{eq:m2}\\
	\delta \xi &=& \frac{\lambda(\gamma-1)}{4!(4\pi)^2}\xi(1+\lambda(\gamma-1))
	\label{eq:xi}
\end{eqnarray}
One can then repeat this procedure {\em ad infinitum}. Apparently, by going to 
infinite order in $\lambda$ we would get the effective non-minimal coupling 
to be
\begin{eqnarray*}
	\xi_{\rm eff} &=& \frac{1}{4!(4\pi)^2}\lambda(\gamma-1)\xi 
	\sum_{n=0}^\infty (\lambda(\gamma-1))^n\\
	&=& \frac{1}{4!(4\pi)^2}\frac{\lambda(\gamma-1)}{1-\lambda(\gamma-1)} \xi
\end{eqnarray*}
corresponding to a finite, multiplicative renormalisation of $\xi$, when 
$1-\lambda(\gamma-1)
\neq 0$, and infinite, multiplicative renormalisation otherwise.\\
One should note that this method is also of use in flat spacetime for strongly 
self-interacting scalar fields.\\
Bunch and Davies, \cite{Bunch}, have found a formula for 
$\langle\phi^2\rangle$ in de Sitter space valid for $m,\xi$ arbitrary:
\begin{displaymath}
	\langle\phi^2\rangle_{BD} = (4\pi\epsilon)^{-2}+\frac{R}{576\pi^2}
	+\frac{m^2+(\xi-\frac{1}{6})R}{16\pi^2}\left(\ln\frac{\epsilon^2R}{12}+
	2\gamma-1+F(\nu)\right)
\end{displaymath}
with $\nu^2=\frac{9}{4}-12\xi-m^2a^{-2}$, $R=12a^2$ and $F(z) = 
\psi(z+\frac{3}{2})+\psi(\frac{3}{2}-z)$, and where $\epsilon$ is a small 
number due to the regularisation procedure chosen. 
The finite part of this differs from ours, by the addition of a constant term, 
due to a different choice of renormalisation. Also, of course, as we go to 
higher order, we have non-linear terms in $R$ and furthermore include the 
derivatives of the curvature scalar.\\ 
For a constant $R$ (or for the derivatives of $R$ negligible) we can
find the mean-field to infinite order in $\lambda$, and get 
\begin{displaymath}
	\langle\phi^2\rangle_{\rm reg}^{(n)} = \frac{\gamma-1}{(4\pi)^2}
	\bar{\tau}_1^{(n)}
	\qquad \bar{\tau}_1^{(n)} = m^2+\xi R+\frac{1}{2}\lambda\langle\phi^2
	\rangle_{\rm reg}^{(n-1)}
\end{displaymath}
leading to
\begin{eqnarray*}
	\langle\phi^2\rangle_{\rm reg}^{(\infty)} &=& \frac{\gamma-1}{(4\pi)^2}\left(
	1+\frac{1}{2}\lambda\frac{\gamma-1}{(4\pi)^2}\left(1+\frac{1}{2}\lambda
	\frac{\gamma-1}{(4\pi)^2}\left(...\right)\right)\right)(m^2+\xi R)\\
	&=& \frac{\gamma-1}{(4\pi)^2}\sum_{n=0}^\infty\left(\frac{1}{2}\lambda
	\frac{\gamma-1}{(4\pi)^2}\right)^n (m^2+\xi R)\\
	&=&\frac{2(\gamma-1)}{2(4\pi)^2-\lambda
	(\gamma-1)}(m^2+\xi R)
\end{eqnarray*} 
which can be seen as a finite renormalisation of $m^2$ and $\xi$.

\subsection{An Example, the Schwarzschild Spacetime}
In this case $\xi R=0$ and the only quantity we need to evaluate is
\begin{equation}
{\cal E}=\frac{1}{4}r^2\cot^2(\theta)-\frac{M}{r^2}-\frac{2M}{r^3}
             -\frac{M^2}{4r^4(1-\frac{2M}{r})}
\end{equation}
where the first term is ignored as an artifact, arising from the 
approximations made; the final result should of course be independent of the 
angles.\footnote{Of course the full heat kernel must be covariant, the apparent
non-covariance -- the $\theta$-dependence -- comes from splitting-up the d'Alembertian
in a non-covariant way. If we were able to carry out the full re-summation this problem
would dissappear by covariance of the procedure, but as we have only a truncated
expression for the heat kernel, we must expect some unphysical effects like this. The
full re-summation would remove the angular-dependence by means of summation formulas
for trigonometric functions. This is likely to give rise to finite term, which we
cannot find by our method at this stage, we will therefore ignore this angular term
altogether. Improved re-summation techniques will then show how accurate this
approximation is, and whether this finite term leads to any testable effect or not.}\\
Since Schwarzschild spacetime is a solution to the vacuum Einstein equations
we get a very simple result for the mean field, namely
\begin{equation}
	\langle\phi^2\rangle_{\rm reg}^{(0)} = \frac{1}{2}(4\pi)^{-2}m^2 
	(1-\gamma) 
\end{equation}
which is independent of $r$.\\
The effective potential then becomes (again taking $V$ to be a $\phi^4$ 
potential) from (19)
\begin{eqnarray}
    V_{\rm eff} &=& \frac{1}{2}m^2\phi_{\rm cl}^2+\frac{1}{4!}\lambda\phi_{\rm cl}^4\nonumber\\
    &&\qquad+\frac{1}{64\pi^2}\left(m^2+\frac{1}{2}\lambda\phi_{\rm cl}^2-
	2Mr^{-2}
    -4Mr^{-3}-\frac{1}{2}M^2r^{-4}(1-\frac{2M}{r})^{-1}\right)^2\times\nonumber\\
    &&\left[\ln\left(m^2+\frac{1}{2}\lambda\phi_{\rm cl}^2-2Mr^{-2}
    -4Mr^{-3}-\frac{1}{2}M^2r^{-4}(1-\frac{2M}{r})^{-1}\right)-\frac{3}{2}\right]\nonumber\\
    &&+\frac{1}{64\pi^2}\Box_0\left(\frac{1}{2}\lambda\phi_{\rm cl}^2-2Mr^{-2}
    -4Mr^{-3}-\frac{1}{2}M^2r^{-4}(1-\frac{2M}{r})^{-1}\right)^{-1}\times\nonumber\\
    &&\left(\partial_p\left(\frac{1}{2}\lambda\phi_{\rm cl}^2-2Mr^{-2}
    -4Mr^{-3}-\frac{1}{2}M^2r^{-4}(1-\frac{2M}{r})^{-1}\right)\right)^2
\end{eqnarray}
from which we get the renormalised mass and couplings to be from (21, 22)
\begin{eqnarray}
    m_{\rm ren}^2 &=& m^2+\frac{\lambda}{32\pi^2}\left(m^2-2Mr^{-2}
    -4Mr^{-3}-\frac{1}{2}M^2r^{-4}(1-\frac{2M}{r})^{-1}\right)\times\nonumber\\
    &&\left[\ln\left(m^2-2Mr^{-2}
    -4Mr^{-3}-\frac{1}{2}M^2r^{-4}(1-\frac{2M}{r})^{-1}\right)-1\right]\nonumber\\
    &&+\frac{\lambda}{96\pi^2}\left(m^2-2Mr^{-2}
    -4Mr^{-3}-\frac{1}{2}M^2r^{-4}(1-\frac{2M}{r})^{-1}\right)^{-2}\times\nonumber\\
    &&\left(\partial_p\left(2Mr^{-2}
    +4Mr^{-3}+\frac{1}{2}M^2r^{-4}(1-\frac{2M}{r})^{-1}\right)\right)^2\\
    \lambda_{\rm ren} &=& \lambda+\frac{3\lambda^2}{32\pi^2}\ln
	\left(m^2-2Mr^{-2}
    -4Mr^{-3}-\frac{1}{2}M^2r^{-4}(1-\frac{2M}{r})^{-1}\right)\nonumber\\
    &&\frac{\lambda^2}{24\pi^2}\left(m^2-2Mr^{-2}
    -4Mr^{-3}-\frac{1}{2}M^2r^{-4}(1-\frac{2M}{r})^{-1}\right)^{-3}\times\nonumber\\
    &&\left(\partial_p\left(2Mr^{-2}
    +4Mr^{-3}+\frac{1}{2}M^2r^{-4}(1-\frac{2M}{r})^{-1}\right)\right)^2
\end{eqnarray}
Far away from the gravitational source, these take on their Minkowski 
spacetime values as
one would expect -- there is no gravitational effect far away. We do, however, 
notice that
the arguments of the logarithms can become negative for certain values of $r$, this then leads
to an imaginary part of the effective action, i.e. to particle production. 
We have plotted the
renormalised mass as a function of the distance to the black hole in figure 1 
with $m=0.0001, \lambda=1, M=1$. Note that the
quantum effects are very much confined to a narrow region around the horizon, 
and very quickly fall
off to their non-gravitational values. Note also that ${\rm Re}~m_{\rm ren}^2
<0$, the renormalised mass also acquires an imaginary part coming from the
logarithm, and it is
\begin{equation}
	{\rm Im}~m_{\rm ren}^2 = \pm i\pi \frac{\lambda}{32\pi^2}
	(m^2-2Mr^{-2}-4Mr^{-3}-\frac{1}{2}M^2r^{-2}(1-\frac{2M}{r})^{-1})
\end{equation}
Here we have used the prescription $\ln (-x) = \ln(x)\pm i\pi$ with $x>0$.\\
It is known, \cite{BD}, that the energy-momentum tensor is divergent on the 
horizon
when one uses Schwarzschild coordinates. This is similar to the divergence 
of the renormalised
mass that we find. Essentially the Schwarzschild coordinates describe an 
external observer, and
the divergence of either $\langle T_{00}\rangle$ or $m_{\rm ren}$ as 
$r\rightarrow 2M$ 
then corresponds to the fact that that an in-falling particle, seen from an 
external observer, never reaches the horizon.

\section{Fermions}
Apparently there is quite a difference between massive fermions, 
described by the Dirac equation, and
massless ones, described by the Weyl equation. The main emphasis in this paper 
is placed on massive particles, and only a few comments on Weyl fermions 
will be made.

\subsection{Free Dirac fermions}
The heat kernel for a Dirac fermion is Clifford algebra-valued and difficult 
to find. We can, however, find the effective potential to one loop order in 
another way. This is done using the following relationship \cite{Casimir}
\begin{equation}
    \zeta_{A^2}(s) = \zeta_A(2s)
\end{equation}
valid for any operator $A$. Since
\begin{equation}
    /\hspace{-3mm}\nabla^2 = (\Box+\xi_fR) 1_4
\end{equation}
it follows that
\begin{equation}
    \zeta_{\nabla\hspace{-2mm}/}(s) = \frac{1}{4} 
      \zeta_{\Box^2+\xi_fR}(\frac{1}{2}s)\label{eq:DiracZeta}
\end{equation}
The factor of one fourth is due to the trace over the Clifford algebra unit 
element.\\
The propagator has been found in \cite{prop} to be
\begin{equation}
    S(x,x') = (i/\hspace{-3mm}\nabla+m)G(x,x')
\end{equation}
where $G(x,x')$ is the propagator for a non-minimally coupled scalar field with
$\xi=\xi_f$. From this we can find the fermion condensate (i.e., mean field) 
by the general relationship
\begin{equation}
    \langle \bar{\psi}\psi\rangle = \left(\lim_{x'\rightarrow x}S(x,x')\right)_{\rm reg}
\end{equation}
where the subscript {\em reg} again indicates a principal value regularisation has to 
be performed to obtain a finite answer. Inserting the explicit formula for 
$G(x,x';\sigma)$, (\ref{eq:heat0}) and integrating over $\sigma$ to get the p
ropagator
for the scalar field, and performing the principal value regularisation we 
arrive at
\begin{eqnarray}
    \langle\bar{\psi}\psi\rangle &=& \frac{1}{2}(4\pi)^{-2}
      \left[i\xi_f(/\hspace{-3mm}\nabla R)
    (1-\gamma)+i(/\hspace{-3mm}\nabla\tau_2)(\xi_fR+m^2)^{-1}\right.\nonumber\\
    &&+i(/\hspace{-3mm}\nabla\tau_3)(\xi_fR+m^2)^{-2}-m(\xi_fR+m^2)
      (\gamma-1)\nonumber\\
    &&+i\xi_f\tau_2(/\hspace{-3mm}\nabla R)(\xi_f R+m^2)^{-2}+2i\tau_2
      (/\hspace{-3mm}\nabla\tau_2)(\xi_fR+m^2)^{-3}\nonumber\\
    &&+6i\tau_2(/\hspace{-3mm}\nabla\tau_3)(\xi_fR+m^2)^{-4}-m\tau_2
      (\xi_fR+m^2)^{-1}\nonumber\\
    &&+2i\xi_f\tau_3(/\hspace{-3mm}\nabla R)(\xi_fR+m^2)^{-3}+6i\tau_3
      (/\hspace{-3mm}\nabla\tau_2)(\xi_fR+m^2)^{-4}\nonumber\\
    &&\left.+24i\tau_3(/\hspace{-3mm}\nabla\tau_3)(\xi_fR+m^2)^{-5}-m\tau_3
      (\xi_fR+m^2)^{-2}\right]
\end{eqnarray}
Notice that for $R$ constant, the mean field of fermions and scalars 
are related by
$\langle\bar{\psi}\psi\rangle_{\rm reg} = m\langle\phi^2\rangle_{\rm reg}$, as
all the terms involving the Dirac operator then vanishes.

\subsection{On Weyl Fermions}
Neutrinos are described by the Weyl equation, i.e., by two-component spinors. 
Given
a Dirac spinor $\psi$ we can form a Weyl spinor by projecting with $\frac{1}{2}
(1\pm\gamma_5)$. Letting $\chi = \frac{1}{2}(1+\gamma_5)\psi$ we get the
following simple equation for $\chi$ from the Dirac equation for $\psi$
\begin{equation}
    \sigma^m(\partial_m+\tilde{\omega}_m)\chi=0
\end{equation}
where $\sigma^0=1_2$ and $\sigma^i,~ i=1,2,3$, are the Pauli matrices. 
This follows by simply writing down
$/\hspace{-2mm}\nabla\frac{1}{2}(1+\gamma_5)$ in the chiral representation of the
Dirac matrices. The new spin connection
$\tilde{\omega}_m$ is related to the old one by
\begin{eqnarray}
    \tilde{\omega}_0 &=& 0\\
    \tilde{\omega}^i &=& i e_0^\mu\omega_\mu^{0i}-e^\mu_0\omega_\mu^{jk}
      \varepsilon_{jk}^{~~~i}
    +ie^\mu_j\omega_\mu^{0k}\varepsilon^{j~i}_{~k}-ie^\mu_j\omega_\mu^{kl}
      \varepsilon_{kl}^{~~~m}
    \varepsilon^{j~i}_{~m}
\end{eqnarray}
A Yang-Mills coupling can be accommodated be simply adding $igA_m^aT_a$ to 
$\tilde{\omega}_m$. Since the Pauli matrices do not form a Clifford algebra proper, 
$\{\sigma_n,\sigma_m\}\neq 2\eta_{nm}$, the square of the Weyl operator will not be 
as simple as that of the Dirac operator, it will in fact still contain a $\sigma^i,
~i=1,2,3$, part. This complication is reminiscent of that of the Dirac operator 
when one includes a non-minimal coupling to the background torsion or to a Yang-Mills
field as shown in the next subsection.\\
We will not pursue the case of Weyl fermions any further in this paper.

\subsection{On Coupling to Yang-Mills Fields and Torsion}
The presence of (background) torsion and gauge field couplings results in a 
change in the expression for the derivative
squared. Explicitly:
\begin{eqnarray}
/\hspace{-3mm} D^2&=&\gamma^mD_m\gamma^nD_n\nonumber\\
&=&\gamma^m\gamma^nD_mD_n+\gamma^m(D_m\gamma^n)D_n\nonumber\\
&=&\eta^{mn}D_mD_n+2i\sigma^{mn}D_mD_n+\gamma^m(D_m\gamma^n)D_n\nonumber\\
&=&\eta^{mn}D_mD_n+i\sigma^{mn}\left[D_m,D_n\right]+\gamma^m(D_m\gamma^n)D_n
\end{eqnarray}
which upon use of the general commutator relation
\begin{equation}
[D_m,D_n]=R_{mn}^{~~~pq}X_{pq}+S_{mn}^{~~~q}D_q+F_{mn}
\end{equation}
(where $R_{mn}^{~~~pq}$ is the Riemann-Christoffel tensor, $S_{mn}^{~~~q}$ 
the torsion and $F_{mn}$ the gauge field strengths and where the generators of
$so_{3,1}$ are denoted by $X_{pq}$ while those of the gauge algebra by $T^a$) becomes
when acting on Dirac spinors
\begin{equation}
/\hspace{-3mm} D^2=\eta^{mn}D_mD_n+i\sigma^{mn}(D^p_{mn}D_p+R_{mn}^{~~~pq}
\sigma_{pq}+F^a_{mn})+\gamma^m(D_m\gamma^n)D_n\\
\end{equation}
which can be brought on the form \cite{Casimir}
\begin{equation}
/\hspace{-3mm} D^2=\Box +\xi_fR+2g\sigma^{pq}F^a_{pq}T_a+g\eta^{pq}A^a_pA^b_q
T_aT_b+{\cal G}(A)+i\sigma^{mn}S^p_{mn}D_p
\end{equation}
From a calculational viewpoint it might seem that the complication due to 
torsion and coupling to gauge fields consists of adding to the operator in 
the heat
kernel equation a first order term which can be removed by the same token as 
it was done for the scalar bosons earlier. However, in the last few steps of the 
determination
of the heat kernel \cite{Casimir} one makes use of the fact that the appearing
quantities commute. This is now not the case. Probably one could remedy the 
situation by commuting anyway, because the commutators would be of next order 
in $\hbar$. In this case it thus is no longer possible to go beyond 1 loop 
order in the gravitational degrees of freedom, except, perhaps, 
through the mean-field approximation.
Furthermore, one should notice that, since we only need $\zeta(s)$, a trace
is performed and thus the problem of non-commutativity may disappear all together
due to the cyclic property of the trace.\\
As we know of no obvious choice of background torsion we will however not 
pursue this line of thought any further.

\subsection{An Example, the Schwarzschild Spacetime}
Since we can express the free fermion effective action in terms of that of a 
non-minimally coupled scalar field (\ref{eq:DiracZeta}), and since $R=0$ for a 
Schwarzschild solution, the effective 
action of a Dirac fermion is simply proportional to that of a scalar field in 
this spacetime. It is probably only for the condensate that we have any chance of 
seeing anything new. In this instance one has the very simple result, 
consistent with the result found for the scalar field,
\begin{eqnarray}
    \langle\bar{\psi}\psi\rangle &=& \frac{1}{2}(4\pi)^{-2}m^3(1-\gamma)
\end{eqnarray}
which is only non-vanishing for Dirac fermions ($m\neq 0$).

\section{Yang-Mills Fields}
The effective potential, $V_{\rm eff}$, 
of a vector boson field is defined from the Euclidean generating
functional by
\begin{equation}
      Z=\int {\cal D}Ae^{-S}\equiv \left(\det\frac{\delta^2S}{\delta A^2}
      \right)^{-1/2}=e^{S_I+S_0}=
      e^{-\frac{1}{2}\zeta'(0)+S_0}
\end{equation}
where $S_0$ is taken to include the ghost and gauge-fixing terms. The ghost 
contribution is $S_{\rm ghost}=-2S_{\rm scalar}$ in the Lorentz gauge; the 
factor of $-2$ is due to the ghosts being independent Grassmann fields even 
though they have
the kinetic action of a minimally coupled scalar field.
Thus the effective potential is related to the zeta function of the field 
operator by the relation (cf. equation (1))
\begin{equation}
    \int V_{\rm eff}(A_{\mu({\rm cl})}^a)\sqrt{g}d^4x = 
    -\frac{1}{2}\zeta_{D_{\mu (b)}^{\nu (a)}}'(0) \label{eq:veff_YM}
\end{equation}
where $D_{\mu (b)}^{\nu (a)}$ is the differential operator of the equations 
of motion
\begin{equation}
    D_{\mu(b)}^{\nu(a)}A_\nu^b = 0
\end{equation}
The equations of motion are to be found from the Lagrangian
\begin{equation}
    {\cal L} = -\frac{g^2}{4}F_{\mu\nu}^aF^{\mu\nu}_a
\end{equation}
where, in curved spacetime, the combined requirements of gauge and Lorentz 
covariance requires that the field strengths be given by
\begin{equation}
    F_{mn}^a = e_m^\mu e_n^\nu(\partial_\mu A_\nu^a-\partial_\nu A_\mu^a
    +igf_{bc}^a A_\mu^b A_\nu^c) 
\end{equation}
see e.g. Ramond 
\cite{Ramond}. This relation is derived by taking the commutator of covariant 
derivatives
\begin{equation}
      D_m=e^\mu_m(\partial_\mu+\frac{i}{2}\omega^{pq}_\mu(x)X_{pq}+
      igA^a_\mu(x)T_a)
    \label{eq:covdiv}
\end{equation}
as in the flat spacetime case (with, again, $X_{pq}$ the generators of the 
Lorentz algebra, i.e.,
$\sigma_{pq}$ when acting on Dirac fermions and so on). With this 
field-strength tensor, the heat kernel equation becomes 
\begin{eqnarray}
&&\hspace{-10mm}\frac{g^2}{4}\left[\delta^a_d\delta^m_r\partial_p\partial^p
         -\delta^a_d(\partial_re^{m\mu}-\partial^me^\mu_r)e^p_\mu\partial_p
         -gf_{d\hspace{3pt}c}^{\hspace{3pt}a}(\partial_rA^{mc}-\partial^mA^c_r)
      \right.\nonumber\\
&&\left.-\frac{1}{2}\delta^m_rg^2f_{edc}f_{f\hspace{3pt}c}^{\hspace{3pt}a}A^e_p
      A^{pf}\right] G_{n(b)}^{r(a)}(x,x';\sigma)=-\frac{\partial}
      {\partial\sigma} G_{n(b)}^{m(a)}(x,x';\sigma)
\end{eqnarray}
which can be written as
\begin{equation}
    D_{r(c)}^{m(a)}G_{n(b)}^{r(c)}(x,x';\bar{\sigma}) =
      -\frac{\partial}{\partial\bar{\sigma}} G_{n(b)}^{m(a)}(x,x';\bar{\sigma})
\end{equation}
where we have introduced $\bar{\sigma}=\frac{g^2}{4}\sigma$.
The gauge indices have been put in parenthesis and will often be suppressed.\\
In analogy to the case of the scalar field the heat kernel then is (see also
\cite{Casimir})
\begin{equation}
    G(x,x';\bar{\sigma}) = G_0(x,x';\bar{\sigma})e^{-{\cal A}\bar{\sigma}
    +\frac{1}{2}{\cal B}\bar{\sigma}^2-\frac{1}{3}{\cal C}\bar{\sigma}^3+...}
      \label{eq:heat_YM}
\end{equation}
where $G$ is a matrix in Lorentz and gauge indices, $G_0$ is the heat kernel of
the operator $\eta^{mn}\partial_m\partial_n$
introduced for the scalar field, and ${\cal A,B,C...}$ are Lie algebra
valued matrices. The zeta function is then given by
\begin{equation}
    \zeta_D(s) = \frac{1}{\Gamma(s)}(4g^{-2})^s\int_0^\infty \bar{\sigma}^{s-1}
      d\bar{\sigma}\int\sqrt{g}d^4x{\rm Tr}~G(x,x;\bar{\sigma})
\end{equation}
where the trace is over as well Lorentz as colour indices. While
the coefficients $\cal A,B,C$ in (\ref{eq:heat_YM}) are given by 
\begin{eqnarray}
    {\cal A}^{m(a)}_{n(b)} &=& \left(\partial_p{\cal E}^{mp}_n+\frac{3}{4}
      {\cal E}^{mp}_k 
    {\cal E}_{np}^k\right)\delta^a_b+gf_{b\hspace{3pt}c}^{\hspace{3pt}a}
      (\partial_nA^{mc}-
      \partial^mA^c_n)+\nonumber\\
    &&\qquad\frac{1}{2}\delta^m_ng^2f_{ebc}f_{d\hspace{3pt}c}^{
      \hspace{3pt}a}A^e_pA^{pd}+\delta^a_bR^m_n
    \label{eq:A}\\
    {\cal B}_{n(b)}^{m(a)} &=& \Box_0{\cal A}_{n(b)}^{m(a)} \label{eq:B}\\
    {\cal C}^{m(a)}_{n(b)} &=& (\partial_p {\cal A}^{m(a)}_{k(c)})(\partial^p 
      {\cal A}^{k(c)}_{n(b)})
      \label{eq:C}
\end{eqnarray}
where
\begin{equation}
    {\cal E}_n^{mp} = \left(\partial_ne^{m\mu}-\partial^me_n^\mu\right) 
      e^p_\mu \label{eq:E}
\end{equation}
Notice that this latter quantity is just the structure coefficients of the
Lie algebra of derivatives $\partial_m$, i.e., $[\partial_m,\partial_n]=
{\cal E}_{mn}^p\partial_p$.\\
We will furthermore use the following approximation to the heat 
kernel\footnote{The matrices $\cal B,C$ contain only derivatives of the 
curvature and of the gauge field, and can thus be considered to be of
a higher order than $\cal A$.} along the diagonal
\begin{equation}
    G(x,x;\bar{\sigma}) \approx (4\pi\bar{\sigma})^{-2}e^{-{\cal A}
      \bar{\sigma}}\left(1+\frac{1}{2}{\cal B}\bar{\sigma}^2
    -\frac{1}{3}{\cal C}\bar{\sigma}^3\right)
\end{equation}
The effective potential which follows from this heat kernel is then
\begin{eqnarray}
    V_{\rm eff}(A) &=&(4\pi)^{-2}{\rm Tr}~\left(-\frac{g^6}{128}{\cal A}^2
      \ln~\frac{g^2}{4}{\cal A} 
    +\frac{3g^6}{256}{\cal A}^2\right.\nonumber\\
    &&\hspace{20mm}\left.-\frac{1}{2}\left(\ln~\frac{g^2}{4}{\cal A}\right)~{\cal 
	B}-\frac{16}{3g^4}{\cal A}^{-1}{\cal C}\right)-
    \frac{g^2}{4}F_{mn}^a F^{mn}_a
\end{eqnarray}
One can get an idea of the physical structure of this effective potential 
by writing
${\cal A}\sim {\cal R}+\partial A+A^2$ with $\cal R$ some curvature term (an 
expression in ${\cal E}_n^{mp}$ and its derivatives, analogous to the standard
expression for the Ricci tensor in terms of the Christoffel symbols and their
derivatives). Then we have
\begin{eqnarray*}
    V_{\rm eff}(A) &\sim & ({\cal R}+\partial A+A^2)^2\left(\ln ({\cal R}+
	\partial A+A^2)+constant\right)\\
    &&-\ln ({\cal R}+\partial A+A^2)\cdot (\Box {\cal R}+\partial^3 A+(\partial A)^2+A\Box A)\\
    &&-({\cal R}+\partial A+A^2)^{-1}(\partial {\cal R}+\partial^2 {\cal R}+A\partial A)^2-F^2
\end{eqnarray*}
which is reminiscent of the $\phi^4$-potential scalar field case. 
The renomalised mass is then found in analogy with the scalar field case
as
\begin{eqnarray}
m^2_{\rm ren}&=&\left.\frac{\partial^2 V_{\rm eff}}{\partial A^p_d \partial A^q_e}
      \right|_{A=0}\delta_{de}\eta^{pq}\nonumber\\
         &=&\delta_{de}\eta^{pq}{\rm Tr}~\left(\left.
      \frac{\partial^2{\cal A}}{\partial A^p_d \partial A^q_e}{\cal A}
    \ln(\frac{g^2}{4}{\cal A})\right.\right.\nonumber\\
          &&\left.\left.+{\cal A}\frac{\partial^2{\cal A}}{\partial A^p_d 
      \partial A^q_e}\ln(\frac{g^2}{4}{\cal A})
              +\frac{1}{2}{\cal A}\frac{\partial^2{\cal A}}{\partial A^p_d 
      \partial A^q_e}
          +\frac{1}{2}{\cal A}^2\frac{\partial^2{\cal A}}{\partial A^p_d 
      \partial A^q_e}{\cal A}^{-1}\right)\right|_{A=0}
\end{eqnarray}
Likewise, the renormalised coupling constant becomes
\begin{eqnarray}
    g_{\rm ren}^2f^{a}_{~bc}f_{ade} &\hspace{-5mm}\equiv &\left.
    -\frac{\partial^4V_{\rm eff}}{\partial A_b^m\partial A_c^n\partial A_d^p
    \partial A_e^q}\right|_{A=0}\eta^{mp}\eta^{nq}\nonumber\\
    &=& g^2f^{a}_{~bc}f_{ade}+\eta^{mp}\eta^{nq}\left[  
      \frac{1}{4}\sum_{perm.}\frac{\partial^2{\cal A}}{\partial A^2}
      \frac{\partial^2{\cal A}}
{\partial A^2}\ln(\frac{g^2}{4}{\cal A})\right.\nonumber\\
  && \hspace{-15mm}+\frac{1}{8}\sum_{perm.}\frac{\partial^2{\cal A}}{\partial A^2}{\cal A}
    \frac{\partial^2{\cal A}}{\partial A^2}{\cal A}^{-1} 
    +\frac{3}{8}\sum_{perm.}\frac{\partial^2{\cal A}}{\partial A^2}
      \frac{\partial^2{\cal A}}{\partial A^2}\nonumber\\
  &&\hspace{-15mm}+\frac{1}{8}{\cal A}\sum_{perm.}\frac{\partial^2{\cal A}}{\partial A^2}
    \frac{\partial^2{\cal A}}{\partial A^2}{\cal A}^{-1}
    +\frac{1}{8}\sum_{perm.}{\cal C}\frac{\partial^2{\cal A}}{\partial A^2}
      {\cal A}^{-1}
    \frac{\partial^2{\cal A}}{\partial A^2}\nonumber\\
 &&\hspace{-15mm}-\frac{1}{2}{\cal A}\left(\frac{\partial^2{\cal A}}{\partial A^n_c 
      \partial A^p_d}{\cal A}^{-1}
    \frac{\partial^2{\cal A}}{\partial A^m_b \partial A^q_e}
     +\frac{\partial^2{\cal A}}{\partial A^m_b \partial A^p_d}{\cal A}^{-1}
    \frac{\partial^2{\cal A}}{\partial A^n_c \partial A^q_e}
    +\frac{\partial^2{\cal A}}{\partial A^m_b \partial A^n_c}{\cal A}^{-1}
    \frac{\partial^2{\cal A}}{\partial A^p_d \partial A^q_e}       
                                             \right)\nonumber\\
&&\hspace{-15mm}
    \frac{1}{2}{\cal A}^2\left(\frac{\partial^2{\cal A}}{\partial A^p_d \partial A^q_e}{\cal A}^{-1}
    \frac{\partial^2{\cal A}}{\partial A^m_b \partial A^n_c}
       +\frac{\partial^2{\cal A}}{\partial A^n_c \partial A^q_e}{\cal A}^{-1}
    \frac{\partial^2{\cal A}}{\partial A^m_b \partial A^p_d}
       +\frac{\partial^2{\cal A}}{\partial A^m_b \partial A^q_e}{\cal A}^{-1}
    \frac{\partial^2{\cal A}}{\partial A^n_c \partial A^p_d}
                                             \right){\cal A}^{-1}\nonumber\\
 &&\hspace{-15mm}+{\cal A}^{-1}\left(\frac{\partial^2{\cal A}}
    {\partial A^n_c \partial A^p_d}{\cal A}^{-1}
    \frac{\partial^2{\cal A}}{\partial A^m_b \partial A^q_e}
     +\frac{\partial^2{\cal A}}{\partial A^m_b \partial A^p_d}{\cal A}^{-1}
    \frac{\partial^2{\cal A}}{\partial A^n_c \partial A^q_e}
     +\frac{\partial^2{\cal A}}{\partial A^m_b \partial A^n_c}{\cal A}^{-1}
    \frac{\partial^2{\cal A}}{\partial A^p_d \partial A^q_e}
                                                     \right)
\left(\Box_0{\cal A}\right)^2\nonumber\\
 &&\hspace{-15mm}+\left(\frac{\partial^2{\cal A}}{\partial A^p_d \partial A^q_e}{\cal A}^{-1}
    \frac{\partial^2{\cal A}}{\partial A^m_b \partial A^n_c}
         +\frac{\partial^2{\cal A}}{\partial A^n_c \partial A^q_e}{\cal A}^{-1}
    \frac{\partial^2{\cal A}}{\partial A^m_b \partial A^p_d}
         +\frac{\partial^2{\cal A}}{\partial A^m_b \partial A^q_e}{\cal A}^{-1}
    \frac{\partial^2{\cal A}}{\partial A^n_c \partial A^p_d}
                                                         \right){\cal A}^{-1}
\left(\Box_0{\cal A}\right)^2\nonumber\\
  &&\hspace{-15mm}\left.\left. +\frac{1}{4}{\cal A}^{-1}(\sum_{perm.}\frac{\partial^2{\cal A}}
           {\partial A^2}{\cal A}^{-1}\frac{\partial^2{\cal A}}{\partial A^2})
           {\cal A}^{-1}(\partial_p {\cal A})(\partial^p {\cal A})\right]\right|_{A=0}
\end{eqnarray}
where we have suppressed gauge indices in the square brackets for clarity. 
Here the `summation over permutations' is to be understood as permutations of 
the order of differentiation subject to the restriction that  
$\frac{\partial}{\partial A^q_e}$ should always appear to the right 
(in the double differentiation one is to differentiate first with respect 
to $A^q_e$, then with respect to $A$ with some
other indices), $\frac{\partial}{\partial A^m_b}$ to the left and 
$\frac{\partial}{\partial A^n_c}$ before $\frac{\partial}{\partial A^p_d}$.
Thus `summation over permutations' is shorthand for six terms in which the 
order of differentiation is permuted, subject to these constraints. 
Furthermore, ${\cal A}$ is given by equation (\ref{eq:A}), ${\cal E}$ is 
given by equation (\ref{eq:E}) and 
\begin{equation}
\left.\frac{\partial^2{\cal A}}{\partial A^h_{\tilde{a}}\partial
A^k_{\tilde{b}}}\right|_{A=0}=\frac{1}{2}g^2f_{eac}f_{d}^{~ac}\eta_{pq}
	(\delta^{e\tilde{b}}\delta^{d\tilde{a}}+
                        \delta^{e\tilde{a}}\delta^{d\tilde{b}}) \equiv g^2\eta_{pq}
	\kappa^{(\tilde{a}\tilde{b})}
\end{equation}
where $\kappa_{ab} = f_{acd}f_b^{~cd}$ and the brackets around the indices 
denote symmetrisation. Note that this quantity only depends on the structure 
of the gauged Lie algebra (more precisely, its Cartan-Killing metric).\\
Also one has
\begin{equation}
\left.\frac{\partial^2}{\partial A^p_d\partial A^q_e}\ln(\frac{g^2}{4}{\cal A})\right|_{A=0}
   =\frac{1}{2}{\cal A}^{-1}\frac{\partial^2{\cal A}}{\partial A^p_d \partial A^q_e}+\frac{1}{2}
    \frac{\partial^2{\cal A}}{\partial A^p_d \partial A^q_e}{\cal A}^{-1}
\end{equation}
and lastly
\begin{eqnarray}
\left.\frac{\partial^4{\cal A}}{\partial A^m_b\partial A^n_c\partial A^p_d\partial A^q_e}\right|_{A=0}
    &=&\frac{1}{2}\left[
                  \frac{\partial {\cal A}^{-1}}{\partial A^n_c \partial A^p_d}
    \frac{\partial {\cal A}}{\partial A^m_b \partial A^q_e}
                  +\frac{\partial {\cal A}^{-1}}{\partial A^m_b \partial A^p_d}
    \frac{\partial {\cal A}}{\partial A^n_c \partial A^q_e}
                  +\frac{\partial {\cal A}^{-1}}{\partial A^m_b \partial A^n_c}
    \frac{\partial {\cal A}}{\partial A^p_d \partial A^q_e}\right.\nonumber\\
   &&\hspace{-10mm}\left.\left.+\frac{\partial {\cal A}}{\partial A^p_d \partial A^q_e}
    \frac{\partial {\cal A}^{-1}}{\partial A^m_b \partial A^n_c}
                  +\frac{\partial {\cal A}}{\partial A^n_c \partial A^q_e}
    \frac{\partial {\cal A}^{-1}}{\partial A^m_b \partial A^p_d}
                  +\frac{\partial {\cal A}}{\partial A^m_b \partial A^q_e}
    \frac{\partial {\cal A}^{-1}}{\partial A^n_c \partial A^p_d}\right]\right|_{A=0}\nonumber\\
\end{eqnarray}
Deriving these results we used `symmetric differentiation', e.g.
\begin{equation}
    \frac{\partial{\cal A}^2}{\partial A^m_b}
      =\frac{\partial {\cal A}}{\partial A^m_b}{\cal A}+{\cal A}
    \frac{\partial {\cal A}}{\partial A^m_b}
\end{equation}
and the relation
\begin{equation}
    \frac{\partial{\cal A}^{-1}}{\partial A^m_b}
      =-{\cal A}^{-1}\frac{\partial{\cal A}}{\partial A^m_b}{\cal A}^{-1}
\end{equation}
which follows from the fact that
\begin{equation}
    0=\frac{\partial}{\partial A^m_b}({\cal A}{\cal A}^{-1})
      =\frac{\partial{\cal A}}{\partial A^m_b}{\cal A}^{-1}
    +{\cal A}\frac{\partial{\cal A}^{-1}}{\partial A^m_b}
\end{equation}
and where we have also used the relation
\begin{equation}
    \frac{\partial}{\partial A^q_e}\ln(\frac{g^2}{4}{\cal A})=
      \frac{1}{2}\left({\cal A}^{-1}
    \frac{\partial {\cal A}}{\partial A^q_e}
                              +\frac{\partial {\cal A}}{\partial A^q_e}
      {\cal A}^{-1}\right)
\end{equation}
which follows when one represent the logarithm by its Taylor expansion 
and then uses `symmetric differentiation' (this
follows if one writes down the expression and performs 
the differentiations component-wise explicitly).\\
Again using ${\cal A}\sim{\cal R}+A+\partial A$ one gets the remarkably 
simple result
($\kappa = {\rm Tr}\kappa_{ab} =\delta^{ab}\kappa_{(ab)}$)
\begin{equation}
	m^2_{\rm ren} = 8g^2\kappa{\rm Tr}\left({\cal R}\left[\ln\frac{g^2}{4}{\cal R}
	+1\right]\right)
\end{equation}
One should note that since $\kappa_{ab}$ is the Cartan-Killing metric, its 
trace is in many cases minus two times the dimension of the Lie algebra. The 
renormalised
mass for the gauge field therefore depends only weakly on the structure of
the gauged Lie algebra and somewhat more strongly on the value of the coupling
constant $g$. The renormalised value of the coupling constant will be
proportional to the trace of the square of $\kappa_{(ab)}$, which potentially
contains more detailed information about the structure of the Lie algebra.\\
Allen has studied scalar QED in Euclidean de Sitter space, i.e. on $S^4$, in 
\cite{Allen}, by means of the $\zeta$-function (the eigenvalues of the 
appropriate 
operators are known in that background). By assumption he has no 
$\lambda^2$-corrections and furthermore $\lambda\sim e^4$, where $e$ is the 
charge
of the scalar field. He finds a formula for the effective action valid for
large radii of the universe which is essential the Coleman-Weinberg potential.
Contrasting with this, even though we do not know the eigenvalues of the 
operators and therefore cannot perform the summation defining $\zeta$ 
explicitly, we do get
$\lambda^2$-corrections as well as a potential valid for small universes too.
As Allen also studied phase transitions we will return to his paper later on.\\
Other results involving gauge bosons have been found by Buchbinder and 
Odintsov,
\cite{BO}, Ishikawa, \cite{Ishikawa}, and Vilenkin, \cite{Vil}, who all find 
the effective action for a scalar field with a $\phi^4$ self-interaction and 
coupled to an electro-magnetic field or, in the case of Buchbinder and 
Odintsov, 
an $SU(2)$ or $SU(5)$ Yang-Mills field possibly with a Yukawa coupling to a 
Dirac fermion too. None of these authors, however, find the effective 
potential for the gauge bosons (nor for the fermions), but only for the
scalar field.

\subsection{Mean Field Approximation and Symmetry Restoration}
In order to find the mean field like we did for scalar fields and fermions, we 
need the matrices $\cal A,B,C$
in the absence of self coupling (i.e. in the limit $g\rightarrow 0$). We then get
\begin{eqnarray}
    \bar{\cal A}^m_n &=& -\partial_p{\cal E}^{mp}_n+\frac{1}{2}{\cal E}^{mp}_k
    {\cal E}_{np}^k +R^m_n\\
    \bar{\cal B}_n^m &=& \Box_0\bar{\cal A}_n^m\\
    \bar{\cal C}^m_n &=& (\partial_p \bar{\cal A}^m_k)(\partial^p \bar{\cal A}^k_n)
\end{eqnarray}
and, analogously to the scalar and fermion case, we find
\begin{eqnarray}
    \langle A_m^a(x) A_n^b(x)\rangle_{\rm reg} &=& 
    -\delta^{ab}\left(\left[\bar{\cal A}
    (\gamma-1)+\frac{1}{2}\bar{\cal B}\bar{\cal A}^{-1}-\frac{1}{3}\bar{\cal C}
    \bar{\cal A}^{-2}\right]\right)_{mn}
    \\
    &=& -\delta^{ab}\left((\bar{\cal A}_{mp}(\gamma-1) +\frac{1}{2}
    \bar{\cal B}_{mk} (\bar{\cal A}^{-1})^k_p-\frac{1}{3}\bar{\cal C}_{mk}
    (\bar{\cal A}^{-1})^k_q(\bar{\cal A}^{-1})^q_p\right)\nonumber\\
\end{eqnarray}
The mean field of course breaks gauge invariance, $\langle A_\mu^k(x)\rangle
\neq 0$, and as pointed out above (and in \cite{EffAct}) this is a 
consequence of virtual particles propagating in the curved spacetime.
Shore \cite{Shore} has pointed out that since $\xi R$ acts as a mass-term for 
a scalar field, and since curvature breaks gauge invariance for the vector 
boson, there is the chance of symmetry {\em restoration} in scalar 
quantum electrodynamics. To 
study this he then considers $S^n$, an Euclidean version of de Sitter space, 
and evaluates the one-loop effective potential for scalar QED by a 
summation of Feynman diagrams (for spacetimes with as much symmetry as $S^n$, 
one can readily write down Feynman rules), a dimensional 
regularisation then shows that this symmetry restoration is indeed possible 
under suitable circumstances. He furthermore finds that minimal couplings 
leads 
to an anomalous mass renormalisation. Unfortunately, since his calculation 
relies heavily on the symmetry of the spacetime $S^n$, his result cannot be 
immediatly extended to other manifolds. But using the framework presented in
this paper, one can now address the question of symmetry 
restoration not just in scalar QED but in a more general Yang-Mills set-up,
at least in 
principle. We will, however, limit ourselves to a few general comments. 
Coupling, 
say, a scalar field with a $\phi^4$ self-interaction to a Yang-Mills field, we 
would to the lowest order just get the effective action for $\phi$ to be the 
corresponding flat spacetime Coleman-Weinberg potential with 
$m^2\rightarrow m^2+\xi R$ and 
$\phi^2\rightarrow \phi^2+\xi R$, as we have mentioned earlier in section 2. 
By fine-tuning the
scalar self-coupling, $\lambda$, its mass $m^2$ and non-minimal coupling 
$\xi$, 
we could then presumably restore gauge symmetry. Our result would, however, 
differ from that of the previously
mentioned authors, \cite{Shore,BO,Vil,Ishikawa}, since we include also the 
gauge 
boson self-interaction using the mean field approach. 
Even the case of a single scalar field without any gauge coupling could 
undergo a symmetry-breaking phase by having its effective mass become 
negative. Similarly, starting out with $m^2<0$, curvature could induce a 
``symmetry-restoration'' ending up with $m_{\rm ren}^2\geq 0$. Similarly, 
adding 
an explicit symmetry-breaking mass-term to the Yang-Mills action, the
curvature could restore the symmetry by leading to a vanishing renormalised 
mass. 
One final thing worth mentioning is that such symmetry-restorations and 
-breakings
induced by curvature could happen, in our approach, even for Ricci flat 
spacetimes ($R_{\mu\nu}=0$) since we still have the $\cal E$-terms.

\subsection{An Example, the Schwarzschild Spacetime}
The explicit expressions for the effective potential of a general Yang-Mills 
field in a  Schwarzschild background are rather involved and will be given bit 
by bit in order to 
understand the various contributions. The essential ingredient is
the matrix ${\cal A}|_{A=0}\equiv {\cal R}$ (a kind of curvature), 
which can be found from (64) to be
\begin{displaymath}
    {\cal R} = \left(\begin{array}{cccc} \frac{M^2}{r^2(r^2-2Mr)} &&&\\
    & \frac{M^2}{r^2(r^2-2Mr)}-2(r^{-2}-2Mr^{-3})&&\\
    && r^{-2}-2Mr^{-3} &\\
    &&& r^{-2}-2Mr^{-3}
    \end{array}\right)
\end{displaymath}
The $A=0$ contribution to the effective potential is then
\begin{eqnarray}
V_{\rm eff}(A=0)&=&-M^2\frac{294912M^2-393216Mr+131072r^2}{3072g^4r^6(r-2M)^2}
\nonumber\\
&&-M^2\frac{32256M^2-43008Mr+15360r^2}{3072r^6(r-2M)
^2}\log(\frac{g^2M^2}{8r^3(r-2M)})\nonumber\\
&&-\frac{294912M^2-196608Mr+32768r^2}{1536g^4r^6}\nonumber\\
&&-\frac{32256M^2-26112Mr+4608r^2}{1536r^6}\log(g^6\frac{r-2M}{8r^3})\nonumber\\
&&-\left\{23887872M^6-60162048M^5r+62652416M^4r^2-34734080M^3r^3\right.\nonumber\\
&&+\left.10878976M^2r^4-
1835008Mr^5+131072r^6\right\}\times\nonumber\\
&&(3072g^4r^6(r-2M)^2(9M^2-8Mr+2r^2))\nonumber\\
&&-\left\{2612736M^6-6912000M^5r+7590912M^4r^2-4426752M^3r^3\right.\nonumber\\
&&\left.+1446912M^2r^4
-251904Mr^5+18432r^6\right\}\times\nonumber\\
&&(3072r^6(r-2M)^2(9M^2-8Mr+2r^2))\log(\frac{g^2(3M-2r)^2}{16r^3(r-M)})
\label{eq:va0}
\end{eqnarray}
This can be seen locally as a quantum correction to the cosmological constant 
and is the only term which is readily written down. One should note that this 
contribution is independent of the Lie algebra one is gauging. The ghost 
contribution is simply given by minus twice the scalar field result and is 
therefore not written down.\\
The renormalised mass is most easily found from (79). This gives
\begin{eqnarray}
m_{\rm ren}^2 &=& 8\kappa g^2\left(2 (r^{-2}-2M r^{-3}) \left(1+\log\left[
	\frac{g^2}{4}(r^{-2}-2M r^{-3})\right]\right)+
	\frac{1+\log\left[\frac{g^2M^2}{4 r^2 (r^2-2Mr)}\right]}{r^2(r^2-2Mr)}
	+\right.\nonumber\\
	&&\left.\left(\frac{M^2}{r^2(r^2-2Mr)}-2(r^{-2}-2Mr^{-3})\right)
	\left(1+\log\left[\frac{g^2}{4}\left(\frac{M^2}{r^2(r^2-2Mr)}
	-2(r^{-2}-2Mr^{-3})\right)\right]\right)\right)	
\nonumber\\
\end{eqnarray}
We have plotted the real part of $m^2_{\rm ren}$ as a function of $g,r$ for
$.001\leq g\leq 1$ and $2.001\leq \frac{r}{M}\leq 5$ and the result is
shown in figure 2. Notice that the renormalised mass vanishes very quickly
away from the horizon, but that it has a narrow region just outside $r=2M$
where ${\rm Re}(m^2_{\rm ren})<0$. The size of this region grows as $g$ grows.
Furthermore, it turns out that the imaginary part becomes negative too
for $r\rightarrow \infty$ for sufficiently large values of the coupling
constant $g$. Thus the vacuum appears to be unstable, at least in this
approximation. The physical 
interpretation of this as particle creation, \cite{GMM}, suggests itself.

\section{Beyond One Loop Order}
Since we now know the effective potentials to one loop order, and since we have
in another paper, \cite{prop}, derived approximate expressions for 
propagators of these
fields in general curved spacetimes (again using the heat kernel), we are now
in a position to go systematically beyond one loop order. In 
fact, the two-loop effective action takes the form, \cite{EffAct}
\begin{eqnarray}
    \Gamma^{(2)} &=& \frac{i}{2(3!)^2}\langle (\frac{\delta^3 S}{\delta\phi^3}
    \phi^3)^2\rangle+
    \frac{1}{4!}\langle \frac{\delta^4S}{\delta\phi^4}\phi^4\rangle
    +\frac{i}{2}\langle(\frac{\delta \Gamma^{(1)}}{\delta\phi}\phi)^2\rangle\\
    &=&-\frac{1}{8}\int dx_1dx_2dx_3dx_4 \frac{\delta^4 S}{\delta \phi(x_1)
    \delta\phi(x_2)\delta\phi(x_3)\delta\phi(x_4)}G(x_1,x_2)G(x_3,x_4)\nonumber\\
    &&-\frac{1}{12}\int dx_1dx_2dx_3dy_1dy_2dy_3 \frac{\delta^3S}{\delta\phi(x_1) 
    \delta\phi(x_2)\delta\phi(x_3)}G(x_1,y_1)G(x_2,y_2)\times\nonumber\\
    &&\hspace{10mm}G(x_3,y_3)\frac{\delta^3S}
    {\delta\phi(y_1)\delta\phi(y_2)\delta\phi(y_3)}
\end{eqnarray}
with $\langle \cdot\rangle$ denoting the expectation value. For a scalar 
field with a $\frac{\lambda}{4!}\phi^4$ coupling as an example 
we can then write this as
\begin{eqnarray}
    \Gamma^{(2)}[\phi] &=&-\frac{1}{8}\lambda\int dx G(x,x)^2
    -\frac{\lambda^2}{48}\int dx dy\phi(x)(G(x,y))^3\phi(y)\\
    &=& -\int dx dy\left(\frac{1}{8}\lambda\langle\phi(x)^2\rangle
    \langle\phi(y)^2\rangle\delta(x,y)+\frac{\lambda^2}{48}
    \phi(x)(G(x,y))^3\phi(y)\right)\nonumber\\
\end{eqnarray}
Into this can then be inserted our expression (\ref{eq:meanfield}) 
for the mean field
and the general formula for the propagator found in \cite{prop}
\begin{displaymath}
    G(x,x') = \alpha_s(x,x') K_1(z)+\beta_s(x,x')K_0(z)
\end{displaymath}
where $s$ refers to the spin, $\alpha_s,\beta_s$ are geometrical quantities, 
$K_i$ are modified Bessel functions and $z=\sqrt{\Delta(x,x')\tau_1(x,x')}$ 
with $\Delta(x,x')$ and $\tau_1(x,x')$ as in section 2. 
In \cite{prop}
approximate expressions was found for the spin-dependent coefficients 
$\alpha_s,\beta_s$, assuming the derivatives of $R$ to be small compared to
$R$ it self (which need not be small).
As the argument $z=\sqrt{\Delta\tau_1}$ of the Bessel functions is a very 
complicated function of
$x,x'$ in the general case, we are not able to write down an analytical
result as explicit as our result for the one-loop term. In any given situation
the integrals in the expression above for $\Gamma^{(2)}$ will have to be
 carried
out either numerically or using some approximation for the $K_i$ or $\Delta$.\\
Another way, of course, is to make use directly of the relation
\begin{equation}
	G(x,x') = -\int_0^\infty G(x,x';\sigma) d\sigma
\end{equation}
relating the propagator to the heat kernel. As mentioned in the section on mean
fields, this
relation follows directly from the spectral decompositions
\begin{eqnarray*}
	G(x,x') &=& \sum_\lambda\lambda^{-1} \bar{\psi}_\lambda(x)
\psi_\lambda(x')\\
	G(x,x';\sigma) &=& \sum_\lambda e^{-\lambda\sigma}\bar{\psi}_\lambda(x)
		\psi_\lambda(x')
\end{eqnarray*}
From this it also follows that (assuming the eigenstates $\psi_\lambda(x)=
\langle x|\lambda\rangle$ to be orthonormal)
\begin{eqnarray}
	G(x,x')^s &=& \sum_\lambda \lambda^{-s}\bar{\psi}_\lambda(x)
	\psi_\lambda(x')\\
	&=& (-1)^{s}\int_0^\infty \sigma^{s-1} G(x,x';\sigma) d\sigma
\end{eqnarray}
and thus (taking the coincidence limit $x'\rightarrow x$)
\begin{eqnarray}
	G(x,x)^2 &=& \int_0^\infty \sigma^2(4\pi\sigma)^{-2} e^{-\tau_1(x)
	\sigma-\tau_2(x)\sigma^2+...} d\sigma\nonumber\\
	&\approx & (4\pi)^{-2}\int_0^\infty e^{-\tau_1(x)\sigma}
	\left[ \sigma^{-1}+\tau_2(x)\sigma+\tau_3(x)\sigma^2\right] 
	d\sigma\nonumber\\
	&=& (4\pi)^{-2}\left[\Gamma(0)+\tau_2\tau_1^{-2}\Gamma(2)
	+\tau_3\tau_1^{-3}\Gamma(3)\right]
\end{eqnarray}
To make this finite we have to replace $\Gamma(0)$ by its principal value 
$-\gamma$, \cite{princip}, and we then get
\begin{equation}
	\int G(x,x)^2dx \approx (4\pi)^{-2}\int\left[ -\gamma
	+\tau_2\tau_1^{-2}+2\tau_3\tau_1^{-3}\right] dx
\end{equation}
Even though this has negative powers of the curvature (remember that 
$\tau_1$ is essentially the curvature scalar) it is still local, this 
contrasts with the second
contribution to $\Gamma^{(2)}[\phi]$ which we get to be (by the same arguments)
\begin{eqnarray}
	\int \phi(x)G(x,y)^3\phi(y) dxdy &\approx & \int \phi(x)
	\left[\Gamma(3) \tau_1(x,y)^{-3}+\Gamma(5)\tau_2(x,y)\tau_1(x,y)^{-5}
	+\right.\nonumber\\
	&&\qquad \left.\Gamma(6)\tau_3(x,y)\tau_1(x,y)^{-6}\right]\phi(y)dxdy
\end{eqnarray}
which is clearly non-local. One should emphasise that there isn't anything 
strange about getting a non-local effective action, in fact quite the 
contrary.\\
Elizalde, Odintsov and Romeo have developed a method of going beyond one-loop 
order for scalar fields
based on a renormalisation group argument, \cite{EOR}, see also Elizalde and 
Odintsov, \cite{EO}. Essentially what they do, is
to write down the most general, local expression
\begin{eqnarray*}
	V_{RG} &=& \frac{1}{4!}\lambda(t) \phi^4(t)-\frac{1}{2}\xi(t)R\phi^2(t)
	+\frac{1}{2}m^2(t)\phi^2(t)+\\
	&&\Lambda(t)+\kappa(t)R+a_1(t)R+a_2(t)C_{\mu\nu\rho\sigma}
	C^{\mu\nu\rho\sigma}+a_3(t){\cal G}
\end{eqnarray*}
where $t$ is the renormalisation group parameter, $t=\frac{1}{2}\ln
\frac{\phi^2}{\mu^2}$, $\Lambda$ is the cosmological constant,
$C_{\mu\nu\rho\sigma}$ is the Weyl conformal tensor, and $\cal G$ is 
the Gauss-Bonnet 
invariant (in four dimensions, $\cal G$ is the sum of squares of the 
Riemann-Christoffel curvature, Ricci tensor and curvature scalar). 
The $t$-derivative of the coefficients are the
respective beta functions found from the renormalisation group calculation. 
This technique is certainly the minimal extension of the usual one-loop 
order result, but one should note the absence of any non-local terms. 
A certain non-locality is to be expected in improvements of the 
Schwinger-DeWitt series, \cite{nonlocal}. 
Such non-locality turns up in the propagators \cite{prop} 
(or more generally, in the heat 
kernel away from the diagonal $x=x'$) 
and will also appear in the final result for $\Gamma^{(n)}$. 
Unfortunately, this non-locality also makes it very difficult to 
actually carry out the calculation 
of, say, the effective action to second-loop order. The RG-improved method of 
Elizalde et al. is certainly much more convenient in this respect, and one 
cannot help feel that somehow a marriage of the two techniques would be 
desirable, but this is, alas, beyond the scope of this paper. 

\section{Quantum Modifications of Classical Critical Points}
We will now turn to a discussion of critical points and phase transitions.
First we study whether a classical critical point survives the transition
to quantum theory (i.e., to the effective potential) or not.
For a scalar field with various self-couplings and non-minimal couplings to
gravity, we show that for a classical critical point to survive the 
quantisation process, the {\em classical} potential must satisfy some extra
conditions.\\
The case of Dirac fermions is again treated by noting the very close 
relationship
with non-minimally coupled scalar fields. Since spinors are 
described by Grassmann numbers the results for scalar fields impose very 
strict conditions on the classical critical point. A few comments on 
non-renormalisable couplings for more than one type of fermion is briefly
discussed.\\
Yang-Mills fields are somewhat more complicated than non-minimally coupled
scalar fields. In any case, in Lorentz gauge we find that a classical critical 
point is only a quantum critical point provided we impose some constraints 
both on the background geometry and upon the Lie algebra we are gauging.\\
The ghost contribution is simply $-2$ times the minimally coupled scalar
field effective action since we are working in Lorentz gauge, and hence will
not be discussed further. We will only discuss critical points in the pure
gauge boson sector.\\

As our starting point we will consider a scalar field with general couplings
$\xi R$ and a general self interaction potential $v(\phi)$, we will 
then relate the spinor and vector case to
this example. We will let $V(\phi)$ include the mass and curvature terms,
i.e., $V(\phi) = \frac{1}{2}m^2\phi^2 + \frac{1}{2}\xi R
\phi^2+v(\phi)$,
$\phi$ is then a critical point of the action if $V=V'=0, V''>0$
when evaluated at $\phi$.\\
Denoting the classical critical point by $\phi_c$ and inserting $V=V'=0$ at
$\phi=\phi_c$ in (\ref{eq:veff}) the very lowest order form for
the effective potential is 
\begin{equation}
	V_{\rm eff}(\phi=\phi_c) = \frac{1}{64\pi^2}(V''(\phi_c)+2{\cal E})^2
	\left[\ln\left(V''(\phi_c)+2{\cal E}\right)-\frac{3}{2}\right]
\end{equation}
similarly
\begin{eqnarray}
	V_{\rm eff}'(\phi=\phi_c) &=& \frac{1}{32\pi^2}(V''(\phi_c)+2{\cal E})
	V'''(\phi_c)\left[\ln\left(V''(\phi_c)+2{\cal E}\right)-\frac{3}{2}
	\right]+\nonumber\\
	&&\frac{1}{64\pi^2}(V''(\phi_c)+2{\cal E})V'''(\phi_c)
\end{eqnarray}
and finally
\begin{eqnarray}
	V_{\rm eff}''(\phi=\phi_c) &=& \frac{1}{32\pi^2}(V'''(\phi_c))^2 
	\left[\ln\left(V''(\phi_c)+2{\cal E}\right)-\frac{3}{2}\right]+
	\nonumber\\
	&&\frac{3}{64\pi^2}(V'''(\phi_c))^2 +\frac{1}{32\pi^2}
	(V''(\phi_c)+2{\cal E})\left[\ln\left(V''(\phi_c)+2{\cal E}\right)-
	\frac{3}{2}\right]V^{(4)}(\phi_c)+\nonumber\\
	&&\frac{1}{64\pi^2}(V''(\phi_c)+2{\cal E})V^{(4)}(\phi_c)
\end{eqnarray}
To the very lowest order in $\hbar$ the requirement that $\phi=\phi_c$ still
be a critical point (i.e., $V_{\rm eff}=V_{\rm eff}'=0, V_{\rm eff}''>0$)
implies
\begin{equation}
	{\cal E} < \frac{1}{2}e^{3/2}\qquad V^{'''}(\phi_c) = 0 
	\qquad V^{(4)}(\phi_c) > 0
\end{equation}
Where the first term, coming from requiring $V_{\rm eff}=0$ (we cannot have 
$V''+2{\cal E}=0$ as this would imply $V'''=V^{(4)}=0$ and hence not be a
critical point), imposes a 
constraint on the background geometry, whereas the remaining conditions only
impose conditions upon $v(\phi)$, the scalar self-interaction potential.
To appreciate the strength of these requirements it is instructive to consider
some particular examples:
\begin{description} 
\item{\bf $\phi^4$-theory:} Let $v=\frac{1}{4!}\lambda\phi^4$, then the 
requirement that the classical critical point $\phi_c\equiv 0$ (with $m^2+\xi 
R>0$, so we get two constraints on the geometry) is still a quantum 
critical point becomes $V^{(4)}=v^{(4)}=\lambda>0$. 
\item{\bf Higgs field:} Consider now the coupling $v=\frac{1}{4!}(\mu-
\sqrt{\lambda}\phi^2)^2$. This has a classical critical point at 
$\phi_c=\pm \sqrt{\frac{3!}{\lambda}(\frac{1}{6}\sqrt{\lambda}\mu
-\xi R)}$ provided 
\begin{displaymath}
	(\sqrt{\lambda}\mu-6\xi R) = 2(\xi R-\sqrt{\lambda}\mu)
     \pm\sqrt{\left(\frac{1}{2}\xi R-\frac{1}{12}\sqrt{\lambda}\mu\right)^2-
     \frac{4}{(4!)^2}\lambda\mu^2}
\end{displaymath} 
which again imposes another constraint upon the background spacetime, namely
(notice the appearance of an imaginary unit)
\begin{displaymath}
	\xi R = \frac{128}{3}\sqrt{\lambda}\mu\left(\frac{13}{576}\pm i
	\frac{\sqrt{139799}}{576}\right)
\end{displaymath}
and
$V'' = \xi R-\frac{1}{6}\sqrt{\lambda}\mu+\frac{1}{2}\lambda\phi_c^2>0$. The
conditions upon quantisation become $V'''=\lambda\phi_c=0$ and $V^{(4)} = 
\lambda>0$. Clearly, this proves that this non-trivial critical point does not
survive either, since $\phi_c=0\Rightarrow V''=0$.
\item{\bf Sine-Gordon theory:} Let now $v(\phi)=\alpha \sin(\beta\phi)$ with
$\alpha,\beta$ some constants. Then the full set of condition reads
\begin{eqnarray*}
	V=0 &\qquad\Rightarrow\qquad& \frac{1}{2}(m^2+\xi R)\phi_c^2 = -
	\alpha \sin(\beta\phi_c)\\
	V'=0 &\qquad\Rightarrow\qquad& (m^2+\xi R)\phi_c=\alpha\beta
	\cos(\beta \phi_c)\\
	V''>0  &\qquad\Rightarrow\qquad& m^2+\xi R-\alpha\beta^2
	\sin(\beta\phi_c)>0\\
	V'''=0 &\qquad\Rightarrow\qquad& \alpha\beta^3\cos(\beta\phi_c)=0\\
	V^{(4)}>0 &\qquad\Rightarrow\qquad& \alpha\beta^4\sin(\beta\phi_c)>0
\end{eqnarray*}
Clearly, this is not possible ($V'''=0$ is only compatible with $m^2+\xi R=0$,
which on the other hand is incompatible with $V^{(4)}>0$), 
hence sine-Gordon theory has no quantum
stable critical points in a curved spacetime. 
\item{\bf Exponential potential:} This is the Liouville-like theory in which
$v(\phi)=\alpha\exp(\beta\phi)$. The requirements read
\begin{eqnarray*}
	V=0 &\qquad\Rightarrow\qquad& \frac{1}{2}(m^2+\xi R)\phi_c^2 =
	-\alpha e^{\beta\phi_c}\\
	V'=0 &\qquad\Rightarrow\qquad & (m^2+\xi R)\phi_c = -\alpha\beta
	e^{\beta\phi_c}\\
	V''>0 &\qquad\Rightarrow\qquad& m^2+\xi R+\alpha\beta^2e^{\beta\phi_c}
	>0\\
	V'''=0 &\qquad\Rightarrow\qquad& \alpha\beta^3 e^{\beta\phi_c}=0\\
	V^{(4)} >0&\qquad\Rightarrow\qquad& \alpha\beta^4 e^{\beta\phi_c}>0
\end{eqnarray*}
Hence this theory cannot have a quantum stable critical point; the 
requirements $V'''=0$ and $V^{(4)}>0$ are mutually incompatible.
\end{description}
This emphasises the importance of taking quantum corrections into account
and not just to rely on experience from classical physics. It furthermore, due
to the extra requirements upon the geometry of the spacetime manifold,
illustrates that using quantum results from flat spacetime
in a curved background scenario, such as the study of inflation in the early 
universe or of black holes, is illegitimate. 

\subsection{Spinor and Vector Fields}
Calculations with Dirac spinors can conveniently, as seen above, be transformed into those
of non-minimally coupled scalar fields by noting
\begin{displaymath}
	(/\hspace{-3mm}\nabla+m)^2 = \Box+m^2+\xi_f R
\end{displaymath}
For a single species of fermions, $\psi^2=0$,
so we cannot have any other interactions than those of the form $\bar{\psi} X
\psi$ where $X$ can be either a scalar or pseudo-scalar field ($X=\phi, X=
\gamma_5\phi$) or a vector or axial vector one ($X=\gamma_m A^m, X=\gamma_5
\gamma_m A^m$). In any case $V'''=V^{(4)}=...=0$, essentially reducing us to
the trivial scalar field case of no self-interactions. If we do have more than
one species of fermions or if the fermion carries colour, then we could have
the non-renormalisable interaction term $v=c_{abcd}\bar{\psi}^a\psi^b 
\bar{\psi}^c\psi^d$ with $c_{abcd}$ a Lorentz scalar. This would bring us
into the $\phi^4$-self coupling case treated above.\\
The case of Yang-Mills fields, on the other hand, provides new insights and
problems. It was proven in a previous section that, in the Lorentz
gauge $\partial_m A^m_a=0$, the effective action becomes (to the very lowest
order)
\begin{equation}
	V_{\rm eff} = (4\pi)^{-2}{\rm Tr}~\left(-\frac{g^6}{128} {\cal A}^2
	\ln\frac{g^2}{4}{\cal A}+\frac{3g^6}{256}{\cal A}^2\right)-V
\end{equation}
with
\begin{eqnarray*}
	V &=&\frac{g^2}{4}F_{mn}^a F^{mn}_a\\
	{\cal A}_{n(b)}^{m(a)} &=& {\cal R}_n^m\delta^a_b +gf_{b~c}^{~a}
	(\partial_n A^{mc}-\partial^mA_n^c)+\frac{1}{2}\delta^m_n g^2
	f_{ebc}f_{d}^{~ac}A_p^eA^{pd}
\end{eqnarray*}
with $\partial_m=e_m^\mu\partial_\mu$ and ${\cal R}_m^n$ some expression in the
derivatives of the vierbeins $e_m^\mu$. The derivatives with respect to the
Yang-Mills field of this effective potential then become
\begin{eqnarray}
	V_{\rm eff}' &=& (4\pi)^{-2}{\rm Tr}~\left(-\frac{g^6}{64}{\cal AA}'
	\ln \frac{g^2}{4}{\cal A}-\frac{g^8}{512}{\cal AA}'+\frac{3g^6}{128}
	{\cal AA}'\right)-V' \label{eq:YMqc1}\\
	V_{\rm eff}''&=& (4\pi)^{-2}{\rm Tr}~\left(-\frac{g^6}{64}({\cal A}')^2
	\ln\frac{g^2}{4}{\cal A}-\frac{g^6}{64}{\cal AA}''\ln\frac{g^2}{4}
	{\cal A}-\frac{3g^8}{512}({\cal A}')^2\right.-\nonumber\\
	&&\left.\frac{g^8}{512}{\cal AA}''+\frac{3g^6}{128}({\cal A}')^2
	+\frac{3g^6}{128}{\cal AA}''\right)-V'' \label{eq:YMqc2}
\end{eqnarray}
where primes denote derivatives with respect to Yang-Mills fields. One should 
note that ${\cal A}''$ is a constant only depending upon the gauged Lie
algebra. One should also notice that gauge invariance is lost, as is often the
case for effective actions, \cite{EffAct}.\\
Now, a necessary condition for $A_m^a$ to be a classical critical point is the
vanishing of the field strength tensor, $F_{mn}^a=0$ which trivially
implies $V=V'=0$. In this case we thus have
\begin{equation}
	\partial_m A_n^a-\partial_n A_m^a = -igf^a_{~bc}A_m^bA_n^c
\end{equation}
For this to be a critical point also after quantisation, we must have ${\cal A}
=0$ in (\ref{eq:YMqc1}) because then $V_{\rm eff}=V_{\rm eff}'=0$. 
Thus we must demand (${\cal R}=\left.{\cal A}\right|_{A=0}$ as before)
\begin{equation}
	{\cal R}_n^m\delta_b^a = ig^2f_{b~c}^{~a}f^c_{~b'c'}A^{b'}_p A^{c'}_n
	\eta^{mp}-\frac{1}{2}\delta^m_n g^2f_{ebc}f_d^{~ac}A_p^eA_q^d\eta^{pq}
\end{equation}
which, using the antisymmetry of $f_{abc}$ in its first two indices, can be
rewritten as
\begin{equation}
	{\cal R}_n^m = -\frac{1}{2}g^2\delta^m_n\kappa_{ab}A^a_pA^b_q\eta^{pq}
\end{equation}
where $\kappa_{ab}= f_{a~e}^{~c}f_{b~c}^{~e}$ is the Cartan-Killing metric on
the Lie algebra and $\eta^{pq}$ is the metric on the tangent space. This 
requirement thus imposes an intricate relationship between the topology of the
manifold (the signature of the tangent space metric), the geometry of the 
manifold (${\cal R}_m^n$ is a kind of curvature formed from the vierbeins) and
the structure of the gauged Lie algebra (through the Cartan-Killing metric).
One immediate consequence is that unless the geometry is such that 
${\cal R}_m^n$ is diagonal, then $A_m^a$ cannot possibly be a quantum
critical point.\\
Considering now the last requirement, $V_{\rm eff}''>0$, we see that we have
to impose furthermore the vanishing of ${\cal A}'$, as we would otherwise
obtain a divergent expression. But
\begin{equation}
	\frac{\partial}{\partial A_p^c}{\cal A}_{n(b)}^{m(a)} =
	\frac{1}{2}\delta_n^mg^2f_{cbe}f_d^{~ae}A_p^d+\frac{1}{2}\delta_n^m g^2
	f_{ebd}f_c^{~ad}A_p^e
\end{equation}
and hence ${\cal A}'=0$ implies the following algebraic constraint upon the
Yang-Mills field
\begin{equation}
	f_{cbe}f_d^{~ae}A_p^d = -f_{ebd}f_c^{~ad}A_p^e
\end{equation}
which only has a solution if $A_m^a=0$ or the Lie algebra structure coefficients 
satisfy
\begin{equation}
	f_{cbe}f_d^{~ae}\delta^{dd'}+f_{ebd}f_c^{~ad}\delta^{ed'}=0
	\qquad \forall a,b,c,d'
\end{equation}
This is a strong requirement to impose on the algebra; $su_2$, for instance,
does {\em not} fulfill it, as can be seen from $\varepsilon_{cb3}
\varepsilon_2^{~13}
+\varepsilon_{2b3}\varepsilon_c^{~13}= -1,0,2$ depending upon the values of
$b,c$. We have not been able to interpret this requirement in the general case;
one feels that it should be related to the cohomology of the Lie algebra, but
no such interpretation springs to mind.
    
\section{Quantum Critical Points}
We have seen that critical points of the classical potential are unlikely to
be critical points of the quantum corrected effective action, so here we want
to see what new critical points turn up in the quantum expression, in order to
find out which new phase transitions could take place due to quantum effects. 
For 
simplicity we will only consider scalar fields and only one concrete example,
namely $\phi^4$-theory.\\
We will denote the quantum critical points by $\phi_{qc}$, these must 
then satisfy
$V_{\rm eff}(\phi_{qc}) = V'_{\rm eff}(\phi_{qc})=0$ and $V_{\rm eff}''
(\phi_{qc})>0$. The first condition implies (assuming $V''+2{\cal E}\neq 0$,
which does not lead to a loss of generality) from equation (97)
\begin{equation}
	\ln (V''+2{\cal E}) -\frac{3}{2} = 64\pi^2\frac{V}{(V''+2{\cal E})^2}
	\label{eq:cond1}
\end{equation}
which we will consider mathematically
as a constraint on the background geometry rather than
as an equation for $\phi_{qc}$, i.e. we will assume in the following that 
the background geometry is such that (\ref{eq:cond1}) is satisfied. 
Inserting this into $V'_{\rm eff}=0$ we get the following non-linear equation
\begin{equation}
	0=V'(V''+2{\cal E})+2VV'''+\frac{1}{64\pi^2}V'''(V''+2{\cal E})^2
	\label{eq:V}
\end{equation}
which is the equation we are going to solve analytically in the $\phi^4$-case. 
Since $V''+2{\cal E}\neq 0$ we can assume it to be positive, in
which case we can write the condition $V_{\rm eff}''>0$, equation(99), as
\begin{equation}
	0<(V''+2{\cal E})(V''+\frac{1}{64\pi^2}V^{(4)})-V'V'''+VV^{(4)}
\end{equation}
where we have also used the above equation (\ref{eq:V}) to simplify matters.
In the case of $V''+2{\cal E}<0$ we just replace the $<$ sign in this equation
by a $>$.\\
Let us now particularise to the case of $v=\frac{1}{4!}\lambda\phi^4$.The
equation we have to solve for $\phi_{qc}$, (\ref{eq:V}), then turns out to
be simply a quadratic equation in $\phi_{qc}^2$
\begin{eqnarray}
	0 &=& \left\{(m^2+\xi R)(m^2+\xi R+2{\cal E})+\frac{\lambda}{64\pi^2}
	(m^2+\xi R+2{\cal E})^2\right\}+\nonumber\\
	&&\phi_{qc}^2\left\{\frac{1}{12}\lambda^2+\frac{5}{3}\lambda m^2
	+\frac{5}{3}\lambda\xi R+\frac{1}{3}\lambda{\cal E}\right\}
	+\frac{\lambda^2}{12}\phi_{qc}^4
\end{eqnarray}
the solution of which evidently is
\begin{eqnarray}
	\phi_{qc} &=& \pm \sqrt{6}\left\{\frac{1}{12}\lambda+\frac{5}{3}
	(m^2+\xi R+\frac{1}{5}{\cal E})\pm\left[\left(\frac{\lambda}{12}+
	\frac{5}{3}(m^2+\xi R+\frac{1}{5}{\cal E})\right)^2-
	\right.\right.\nonumber\\
	&&\qquad\left.\left. \frac{1}{3}\left((m^2+\xi R)(m^2+\xi R+2{\cal E})
	+\frac{\lambda}{64\pi^2}(m^2+\xi R+2{\cal E})^2\right)\right]^{1/2}
	\right\}^{1/2}\nonumber\\
\end{eqnarray}
This can then be inserted into the two remaining conditions, $V_{\rm eff}=0$
and $V_{\rm eff}''>0$ to give constraints linking $m^2,\lambda$ to $\xi R,
{\cal E}$. In order to avoid too complicated expressions, we will assume $m=
\xi=0$ for now. This should bring out the underlying physics more clearly. The
equation $V_{\rm eff}=0$ then reads
\begin{eqnarray}
0&=&\left( 2{\cal E} + 3\lambda\left( \frac{{\cal E}}{3} + 
	\frac{\lambda}{12} + 
         \sqrt{\left( \frac{{\cal E}}{3} + \frac{\lambda}{12} \right)^
                    2 \mp \frac{{\cal E}^2\lambda}{48\pi^2}} 
\right)  \right)^2 \nonumber\\
&&\times
     \left( -\frac{3}{2} + \log \left[2{\cal E} + 
         3\lambda\left( \frac{{\cal E}}{3} + \frac{\lambda}{12} + 
        \sqrt{\left( \frac{{\cal E}}{3} + \frac{\lambda}{12} \right)^2 
\mp 
        \frac{{\cal E}^2\lambda}{48\pi^2}}\right) \right] \right) 
\end{eqnarray}
This gives us the two quartic equations for $\lambda$ as a function of $\cal E$
\begin{eqnarray}
	0&=& \lambda^4 \frac{155}{164}+
	\lambda^3(\frac{143}{18}
	{\cal E}\pm \frac{1}{48 \pi^2}{\cal E}^2)+16{\cal E}\lambda^2
	+4{\cal E}^2\lambda+4{\cal E}^2\\
	0 &=& \lambda^4\frac{155}{164}\pm 
	\frac{3{\cal E}^2}{16\pi^2}\lambda^3+\lambda^2(\frac{3}{2}{\cal E}-
	\frac{1}{2}e^{3/2})+\lambda(6{\cal E}^2-2{\cal E}e^{3/2})
	+(e^3+9{\cal E}^2-6e^{3/2}{\cal E})\nonumber\\
\end{eqnarray}
Hence, for each choice of sign, $\pm$, there are at most eight solutions, 
bringing the total maximum number of values for $\lambda=\lambda_{qc}$ up 
to sixteen. It 
turns out that these, whilst easy to find, are very complicated functions of 
$\cal E$. Since no interesting physics seems to be contained in their exact 
form, we will not list the solutions.\\
The positivity requirement, $V_{\rm eff}''>0$, turns out to be
\begin{eqnarray}
0&<&\frac{1}{2}\lambda\left( 2{\cal E} + 3\lambda\left(\frac{{\cal E}}{3} + 
\frac{\lambda}{12} + 
          \sqrt{\left(\frac{{\cal E}}{3}+\frac{\lambda}{12} \right)^2\mp 
          \frac{{\cal E}^2\lambda}{48\pi^2}} \right)  \right) +\nonumber\\
&& 6\lambda^2\left( \frac{{\cal E}}{3} + \frac{\lambda}{12} + 
     \sqrt{\left( \frac{{\cal E}}{3} + \frac{\lambda}{12} \right)^2 \pm 
          \frac{{\cal E}^2\lambda}{48\pi^2}} \right)\times\nonumber\\
&& 
   \left( -\frac{3}{2} + \log \left[2{\cal E} + 
       3\lambda\left( \frac{{\cal E}}{3} + \frac{\lambda}{12} + 
          \sqrt{\left( \frac{{\cal E}}{3} + \frac{\lambda}{12} 
	\right)^2 \mp
       \frac{{\cal E}^2\lambda}{48\pi^2}} \right) \right] \right) 
\end{eqnarray}
One can then insert one of the values for $\lambda_{qc}$ in order to get a
constraint upon $\cal E$, and thus upon the background geometry (which
must of course be consistent with the constraint imposed on $\cal E$ from
(110) above). Figure 3 is a 
surface plot of the right hand side of this inequality as a function of
$\lambda, \cal E$ in the region $0.001\leq\lambda\leq 1, -5\leq {\cal E}\leq 
5$. The real part of $V_{\rm eff}''$ turn out to be the same
for either choice of sign, but for ${\cal E}<0$, $V_{\rm eff}''$ acquires
an imaginary aprt (different from the two choices of sign) for sufficiently
large values of $\lambda$, hence signaling a violation of the above condition
in that regime. For ${\cal E}>0$ the condition is generally satisfied.

\section{The Energy-Momentum Tensor}
For the study of back-reaction of the matter degrees of freedom upon 
the geometry
one must obtain a formula for the renormalised energy-momentum tensor. Thus to 
know this quantity is essential for cosmological applications in which one 
takes the evolution of the universe due to the quantum fields present in it 
into account. Of importance is this respect is various mathematical features 
such as the possibility of violation of energy conditions (the Casimir energy 
density can be arbitrarily large and negative,
contrary to what is possible classically) and the conformal anomaly 
which shows the quantum break-down of the conformal invariance present 
in the classical theory. 
These questions are the subject of the following section.\\
The Einstein equations are $G_{\mu\nu}=T_{\mu\nu}$, where the tensor 
$G_{\mu\nu}$ is found by varying the Einstein-Hilbert action,
\begin{displaymath}
    G_{\mu\nu} =2g^{-1/2}\frac{\delta}{\delta g^{\mu\nu}} \int R\sqrt{g}d^4x = 
	R_{\mu\nu}-\frac{1}{2}g_{\mu\nu}R
\end{displaymath} 
The classical equations of motion are therefore
\begin{displaymath}
	0=\frac{\delta}{\delta g^{\mu\nu}}(S_{\rm EH}+S_{\rm matter})
\end{displaymath}
Or in other words
\begin{equation}
	T_{\mu\nu} = 2g^{-1/2}\frac{\delta}{\delta g^{\mu\nu}}S_{\rm matter}
\end{equation}
Given the effective action $\Gamma$, we can then find the renormalised 
energy-momentum tensor by, \cite{BD,EffAct}
\begin{equation}
    \langle T_{\mu\nu}\rangle = 2 g^{-1/2}\frac{\delta \Gamma}{\delta g^{\mu\nu}}
\end{equation}
where $\Gamma = \int V_{\rm eff} \sqrt{g}d^4x$.\\
Looking at the explicit form, (\ref{eq:veff}), for the effective potential 
for a non-minimally coupled scalar field we see that in order to
find $\langle T_{\mu\nu}\rangle$ all we need to know then is
\begin{displaymath}
    \frac{\delta}{\delta g^{\mu\nu}} \int {\cal E}\sqrt{g}d^4x \equiv 
	\frac{1}{2}\sqrt{g}H_{\mu\nu}
\end{displaymath}
In terms of $G_{\mu\nu},H_{\mu\nu}$ the renormalised energy-momentum tensor 
becomes
(by simply carrying out the variation with respect to the metric for a 
non-minimally coupled scalar field)
\begin{eqnarray}
    \langle T_{\mu\nu}\rangle_{\rm scalar} 
	&=& 2 g g_{\mu\nu} (\frac{1}{2}m^2\phi_{\rm cl}^2
    +V(\phi_{\rm cl}))+\frac{1}{32\pi^2}(m^2+V''(\phi_{\rm cl})+\xi R+2{\cal E})
    \nonumber\\
    &&\times\left[\ln\left(m^2+V''+\xi R+2{\cal E}\right)-1\right]
    \left( \xi G_{\mu\nu}+2H_{\mu\nu}\right)\nonumber\\
    &&
    +\frac{1}{64\pi^2}\Box_0\left(\xi G_{\mu\nu}+H_{\mu\nu}\right)\nonumber\\
    &&+\frac{1}{96\pi^2}\left(m^2+V''+\xi R+{\cal E}\right)^{-2}
    \left(\partial_p\left(V''+\xi R+{\cal E}\right)\right)^2
    \left(\xi G_{\mu\nu}+H_{\mu\nu}\right)\nonumber\\
    &&+\frac{1}{96\pi^2}\partial^p\left[\left(m^2+V''+\xi R+{\cal E}\right)^{-1}
    \partial_p\left(V''+\xi R+{\cal E}\right)\right]
    \left(\xi G_{\mu\nu}+H_{\mu\nu}\right)\nonumber\\
	\label{eq:T}
\end{eqnarray}
In order to find an explicit expression for the tensor $H_{\mu\nu}$ (which, 
like the
Einstein tensor, is a geometrical object) it is convenient to vary with respect
to the vierbein, instead of the metric.  Using (see Ramond \cite{Ramond})
\begin{equation}
    \delta e = e e^m_\mu\delta e_m^\mu
\end{equation}
we get
\begin{eqnarray}
    \frac{\delta}{\delta e_\nu^n}\int{\cal E}\sqrt{g}d^4x &=& 
    \frac{1}{2}\partial^\nu(e^{-1}\partial_\mu(ee_n^\mu))
    +\frac{1}{2}e^{-1}e_n^\nu (\partial^\rho e_\rho^m) \partial_\sigma
    (ee^\sigma_m)\nonumber\\
    &&+\frac{1}{2}e^{-1}e^\nu_n\partial_\mu(e^m_\sigma\partial^\sigma e) e^\mu_m
    -\frac{1}{2}e^{-1}\partial_\mu(e_\rho^p\partial^\rho e)e^\mu_pe^\nu_n\nonumber\\
    &&-\frac{1}{2}e^{-2}e^\nu_n\eta^{pq}\partial_\mu(ee^\tau_p)\partial_\rho(ee^\sigma_q)
    g^{\mu\rho}g_{\sigma\tau}\nonumber\\
    &&-\frac{1}{4}ee^\nu_ne^\tau_p\partial_\mu(e^{-2}\partial_\rho(ee^\sigma_m)
    g^{\mu\rho}g_{\sigma\tau})\eta^{mp}\nonumber\\
    &&-\frac{1}{4}e\partial_\mu(e^{-2}\partial_\rho(ee^\sigma_m)g^{\mu\rho}\delta^\nu_\sigma
    \delta^m_n)\nonumber\\
    &&-\frac{1}{4}ee^\nu_ne^\tau_p\partial_\rho(e^{-2}\partial_\mu(ee^\sigma_m)
    g^{\mu\rho}g_{\sigma\tau})\eta^{mp}\nonumber\\
    &&-\frac{1}{4}e\partial_\rho(e^{-2}\partial_\mu(ee^\sigma_m)g^{\mu\rho}\delta^\nu_\sigma
    \delta^m_n)\nonumber\\
    &&-\frac{1}{4}e^{-2}\partial_\mu(ee^\tau_p)\partial_\rho(ee^\sigma_m)
    g_{\sigma\tau}(e^\mu_n g^{\rho\nu}+e^\rho_ng^{\mu\nu})\nonumber\\
    &&+\frac{1}{4}e^{-2}\partial_\mu(ee^\tau_p)\partial_\rho(ee^\sigma_m)
    g^{\mu\rho}(\delta^\nu_\tau \eta_{pn}e^p_\sigma+\delta^\nu_\sigma
    \eta_{pn}e_\tau^p)\nonumber\\
    &&+{\cal E}e^\nu_n
\end{eqnarray}
Using the relationship between the vierbein and the metric
\begin{eqnarray}
    \delta g_{\mu\nu} &=& (\delta e_\mu ^m e_\nu ^n+e_\mu^m \delta e_\nu^n)
	\eta_{mn}\label{eq:deltag}\\
    \delta e^\mu_n &=& -e^\mu_n e_m^\nu\delta e_\nu^m
\end{eqnarray}
which follows from the defining relations, $g_{\mu\nu}=e_\mu^me_\nu^n
\eta_{mn}$ 
and $e_\mu^m e^\mu_n=\delta^m_n$, for the vierbein directly. We then get
\begin{equation}
    H_{\mu\nu} \equiv 2g^{-1/2}\frac{\delta}{\delta g^{\mu\nu}}\int{\cal E}\sqrt{g}d^4x
    = 2g^{-1/2}\frac{\delta g^{\mu\nu}}{\delta e^\rho_n}\frac{\delta}{\delta g^{\mu\nu}}
    \int {\cal E}\sqrt{g}d^4x = g^{-1/2}g_{\mu\nu}e_n^\rho\frac{\delta}{\delta e_n^\rho}
    \int{\cal E}\sqrt{g}d^4x
\end{equation}
Note that
\begin{equation}
	H_{\mu\nu} = \frac{1}{4}g_{\mu\nu} H \qquad H \equiv g^{\mu\nu}H_{\mu\nu}
	= \frac{1}{4}g^{-1/2}e^\rho_n\frac{\delta}{\delta e^\rho_n}\int {\cal E}
	\sqrt{g}d^4x
\end{equation}

For Dirac fields, we have seen that $\zeta_A(s) = \zeta_{A^2}(\frac{1}{2}s)$ 
and that the square of the Dirac operator was just the Klein-Gordon operator 
with a 
non-minimal coupling $\xi=\xi_f$. Therefore the renormalised energy-momentum 
tensor for such spinor
fields is essentially the same as for non-minimally coupled scalar fields 
(up to multiplicative constants).\\

Yang-Mills fields on the other hand, gives rise to something
new. In order to carry out the variation of $V_{\rm eff}$ with respect to the 
metric (or equivalently, the vierbein) we then see from 
(\ref{eq:veff_YM}) that 
we need to evaluate the variation of the tensor ${\cal E}^{mp}_n$ and of the 
matrix-valued coefficient ${\cal A}^{m(a)}_{n(b)}$. Using their definitions, 
(\ref{eq:A},\ref{eq:E}), we then get
\begin{eqnarray}
	\frac{\delta{\cal E}^{mp}_n}{\delta e^q_\rho} &=& -e_q^\rho{\cal E}^{mp}_n
	+\delta^p_q\eta^{km}\left(e^\nu_n\partial_\nu e_k^\rho-e^\nu_k\partial_\nu
	e^\rho_n\right)+\nonumber\\
	&&\eta^{km}e^\rho_q\left(e^\mu_k\partial_\nu(e_n^\nu e^p_\mu) -e^\mu_n
	\partial_\nu(e^\nu_k e^p_\mu)\right)
\end{eqnarray}
and
\begin{eqnarray}
	\frac{\delta {\cal A}_{n(b)}^{m(a)}}{\delta e^q_\rho} &=& 
	\left[ -e^\mu_p e^\rho_q
	\partial_\mu{\cal E}_n^{mp}+e^\mu_p\partial_\mu\frac{\delta{\cal E}_n^{mp}}
	{\delta e_\rho^q}+\frac{3}{2}\frac{\delta{\cal E}_k^{mp}}{\delta e^q_\rho}
	{\cal E}_{np}^k\right]\delta_b^a+\nonumber\\
	&&gf_{b~c}^{~a}\left[-e^\mu_ne^\rho_q\partial_\mu(e_\nu^m a^{\nu c})
	-\partial_\mu(e^\mu_nA^{\nu c})\delta^m_q\delta^\rho_\nu-\right.\nonumber\\
	&&\left.\delta^m_q\partial^\rho(e^\nu_nA_\nu^c)-
	\partial^\mu(e^m_\mu A_\nu^c)e^\nu_ne^\rho_q\right]-\nonumber\\
	&&g^2\delta^m_n f_{eb}^{~~c}f_{d~c}^{~a}e^\mu_p e^\rho_qe^\nu_l A_\mu^e
	A_\nu^d\eta^{pl} +\delta^a_b\frac{\delta R^m_n}{\delta e^q_\rho}
\end{eqnarray}
The variation of the remaining coefficients ${\cal B,C}$ in the effective
action for Yang-Mills fields can then be expressed in terms of the variation of
${\cal A}$. Explicitly
\begin{eqnarray}
	\frac{\delta {\cal B}^{m(a)}_{n(b)}}{\delta e^q_\rho} &=& -e^\rho_q 
	{\cal B}^{m(a)}_{n(b)}+\eta^{kl}(\partial_\mu e_k^\mu)e^\nu_le^\rho_q
	\partial_\nu
	{\cal A}_{n(b)}^{m(a)}+\Box_0\frac{\delta{\cal A}^{m(a)}_{n(b)}}
	{\delta e^q_\rho}
	\\
	\frac{\delta{\cal C}^{m(a)}_{n(b)}}{\delta e^q_\rho} &=& 
	-2e^q_\rho{\cal C}^{m(a)}
	_{n(b)}-2\eta^{pl}\partial_\mu(e^\mu_pe^\nu_l\partial_\nu
	{\cal A}^{k(c)}_{n(b)})
	\frac{\partial{\cal A}^{m(a)}_{k(c)}}{\delta e^q_\rho}
\end{eqnarray}
where we have used that $\Box_0\equiv \eta^{pq}\partial_p
\partial_q = \eta^{pq}e^\mu_p\partial_\mu(e^\nu_q\partial_\nu)$ hence it gives a
contribution to the variation. From
\begin{equation}
	\frac{\delta\Gamma}{\delta g^{\mu\nu}} = -2g_{\mu\nu}e^q_\rho
	\frac{\delta\Gamma}{\delta e^q_\rho}
\end{equation}
which is simply the chain rule for functional derivation together with 
(\ref{eq:deltag}), we get the following formula
for the renormalised energy-momentum tensor for a Yang-Mills field (with $A_\mu^a
\neq \langle A_\mu^a\rangle$, the curvature-induced mean field)
\begin{eqnarray}
	\langle T_{\mu\nu}\rangle_{\rm YM} 
	&=& T_{\mu\nu}^{\rm cl}-g^{-1/2}g_{\mu\nu} e^q_\rho
	(4\pi)^{-2}{\rm Tr}\left(-\frac{g^6}{128}\left\{{\cal A},
	\frac{\delta {\cal A}}
	{\delta e^q_\rho}\right\}\ln ~\frac{g^2}{4}{\cal A}-\right.\nonumber\\
	&&\frac{g^8}{1024}{\cal A}^2\left\{{\cal A}^{-1},\frac{\delta{\cal A}}
	{\delta e^q_\rho}\right\}+\frac{3g^6}{128}{\cal A}\frac{\delta{\cal A}}
	{\delta e^q_\rho}-\nonumber\\
	&&\frac{g^2}{16}\left\{{\cal A}^{-1},\frac{\delta{\cal A}}
	{\delta a^q_\rho}\right\}
	{\cal B}-\frac{1}{2}(\ln~\frac{g^2}{4}{\cal A})\left(-e^\rho_q 
	{\cal B}+\eta^{kl}(\partial_\mu e_k^\mu)e^\nu_le^\rho_q\partial_\nu
	{\cal A}+\Box_0\frac{\delta{\cal A}}{\delta e^q_\rho}\right)
		+\nonumber\\
	&&\left.\frac{16}{3g^4}\left\{{\cal A}^{-2},\frac{\delta{\cal A}}
	{\delta e^q_\rho}\right\}
	{\cal C}-\frac{16}{3g^4}{\cal A}^{-1}\left(-2e^q_\rho{\cal C}
	-2\partial_\mu(g^{\mu\nu}\partial_\nu{\cal A})
	\frac{\partial{\cal A}}{\delta e^q_\rho}\right)\right) \label{eq:T_YM}
\end{eqnarray}
where $T_{\mu\nu}^{\rm cl}$ is the classical expression for the energy-momentum 
tensor. Even with $A_\mu^a$ a classical external field, and not the curvature 
dependent mean field, this is a highly complicated function of the geometric 
variables and we can only say very little about it.

\subsection{The Conformal Anomaly}
From (\ref{eq:T}) we can also calculate the conformal anomaly, i.e., the 
renormalised value of the trace of the energy-momentum tensor. For a scalar 
field we get
\begin{eqnarray}
	\langle T\rangle &=& 8g(\frac{1}{2}m^2\phi_{\rm cl}^2+V(\phi_{\rm cl}))
	+\frac{1}{32\pi^2}(m^2+V''+\xi R+2{\cal E})\times\nonumber\\
	&&\left[\ln(m^2+V''+\xi R+2{\cal E})-1\right](2H-\xi R)+\nonumber\\
	&&\frac{1}{64\pi^2}g^{\mu\nu}\Box_0(\xi G_{\mu\nu}+H_{\mu\nu})+\nonumber\\
	&&\frac{1}{96\pi^2}(m^2+V''+\xi R+{\cal E})^{-2}(\partial_p(V''+\xi R+
	{\cal E}))^2(H-\xi R)+\nonumber\\
	&&\frac{1}{96\pi^2}\partial^p\left[(m^2+V''+\xi R+{\cal E})^{-1}
	\partial_p(V''+\xi R+{\cal E})\right](H-\xi R)
\end{eqnarray}
Whenever this is non-vanishing, conformal symmetry is broken even for $m=0,\xi=
\frac{1}{6}$. Notice that in spacetimes with $H=\xi R$ the trace of 
the energy-momentum tensor simplifies, and all the terms 
involving derivatives of the curvature (except the $\Box_0$-term) vanish.\\ 
This result depends on $\phi_{\rm cl}^2$, i.e.
on the particular classical configuration. By interpreting $\phi_{\rm cl}^2$ as
$\langle \phi^2\rangle$, the above formula tells us how curvature can induce a 
breaking of conformal invariance non-perturbatively, since $\langle 
\phi^2\rangle$ is a function of the curvature only.\\

Upto numerical coefficients this result also holds for spinors as follows once 
again from the relationship between their zeta functions. \\

For Yang-Mills fields, however, we get
something different.  Let us first note that $\langle T_{\mu\nu}\rangle$, 
(\ref{eq:T_YM}), in this case can be written as
\begin{equation}
	\langle T_{\mu\nu}\rangle_{\rm YM} = T_{\mu\nu}^{\rm cl} - 
	g^{-1/2}g_{\mu\nu} Y
\end{equation}
where $Y$ is given by the vierbein times the trace in (\ref{eq:T_YM}). 
Thus the quantum corrections to the energy-momentum tensor are proportional to 
the metric. From this it follows that the conformal anomaly for 
Yang-Mills fields is simply 
\begin{equation}
	\langle T\rangle_{\rm YM} = -4g^{-1/2}Y
\end{equation}
The vanishing or non-vanishing of $Y$ is a very complicated 
condition depending on as well the background geometry as on the structure 
of the gauged Lie algebra. At present, we are unable to say anything 
more about it.

\subsection{Violation of Energy Conditions}
The structure of the energy-momentum tensor is important for a number of 
reasons not only for the possibility of breaking conformal invariance as 
the conformal anomaly calculated above shows, but also in relation 
with the various singularity theorems. 
The singularity theorems of Penrose and Hawking and Penrose, respectively, 
rely on 
the weak and the strong energy conditions, whereas the second law of black hole
thermodynamics is based on the non-violation of the null energy condition, and
the topological censorship theorem on that of the averaged null condition. And 
finally, the cosmic censorship conjecture is based on the dominant energy 
condition. These conditions, \cite{Visser,GMM}, are listed in table 1. \\

Start with a scalar field, the energy-momentum tensor being given 
by (\ref{eq:T}). Let us now consider an 
arbitrary null vector, then $H_{\mu\nu} k^\mu k^\nu = \frac{1}{4}Hk^2 = 0$, 
thus the $H_{\mu\nu}$
tensor terms disappear from $\langle T_{\mu\nu}\rangle k^\mu k^\nu$ and we are
left with
\begin{equation}
	\langle T_{\mu\nu}\rangle k^\mu k^\nu = \frac{\xi}{32\pi^2}(G_{\mu\nu}
	k^\mu k^\nu)X(g,\phi)+\frac{\xi}{96\pi^2}\Box_0 G_{\mu\nu}k^\mu k^\nu
\end{equation}
with
\begin{eqnarray}
	X(g,\phi) &=& (m^2+V''+\xi R+2{\cal E})\left[\ln(m^2+V''+\xi R+
	2{\cal E})-1\right]+\nonumber\\
	&&\frac{1}{3}(m^2+V''+\xi R+{\cal E})^{-2}(\partial_p(V''+\xi R+
	{\cal E}))^2+\nonumber\\
	&&\frac{1}{3}\partial_p\left[(m^2+V''+\xi R+{\cal E})^{-1}\partial^p
	(V''+\xi R+{\cal E})\right]
\end{eqnarray}
Now, by the Einstein equations, $G_{\mu\nu} = T_{\mu\nu}^{\rm cl}$, the 
classical energy-momentum tensor and we know, \cite{Visser}, that a 
classical scalar field
only violates SEC, i.e., $G_{\mu\nu} k^\mu k^\nu \geq 0$. Thus NEC is violated
if and only if $X < 0$. To the lowest order this condition reads
\begin{equation}
	(m^2+\frac{1}{2}\lambda\phi^2+\xi R+2{\cal E}) \left[\ln (m^2+\frac{1}{2}
	\lambda\phi^2+\xi R+2{\cal E})-1\right] < 0 \label{eq:X}
\end{equation}
which is clearly possible {\em a priori}.\\
It is quite normal for a Casimir energy
to violate all the non-averaged energy conditions, and this is then also 
what we see
here. This implies the breakdown of the singularity theorems in the presence of
quantum matter fields. To see whether the averaged conditions are 
violated or not,
we would need to consider a specific spacetime and a specific geodesic, which 
we will not do here.\\
Requiring the logarithm in (\ref{eq:X}) to be real, leads to (\ref{eq:X})
being reformulated as
\begin{equation}
	0 < m^2+\frac{1}{2}\lambda\phi^2+\xi R+2{\cal E} < e
\end{equation}
For a Ricci-flat spacetime, such as the Schwarzschild solution for instance, 
this implies
the following bound on the value of the scalar field (if NEC is violated)
\begin{equation}
	\frac{1}{2}\lambda\phi^2 < e-m^2-2{\cal E}
\end{equation}
and we clearly have the possibility of this being true at the horizon, 
signalling 
the possible break down, locally in the vicinity of the horizon, of black hole 
thermodynamics due to non-pertubative quantum effects.\\

Dirac fields differ from scalar fields in that $\xi=\xi_f$ and in an over 
all factor 
of $-2$ (the minus being due to the Grassmannian nature of spinors and the 
factor 
two being the number of spin degrees of freedom), one should therefore just 
reverse all signs in order
to get a condition for violation of NEC for Dirac fermions.\\

For Yang-Mills fields it is once more convenient to write
\begin{equation}
	\langle T_{\mu\nu}\rangle_{\rm YM} = T_{\mu\nu}^{\rm cl} - 
	g^{-1/2}g_{\mu\nu} Y
\end{equation}
with $Y$ being a trace (see (\ref{eq:T_YM}) for explicit details). It then 
follows
that Yang-Mills fields {\em cannot} violate NEC even with quantum 
corrections taken 
into account, since these are proportional to $g_{\mu\nu}$, thus $\langle 
T_{\mu\nu}\rangle_{\rm YM}k^\mu k^\nu \equiv 0$.

\section{Discussion and Conclusion}
The approach chosen in this paper differs from that by other authors, as
the chosen expansion of the heat
kernel is not just an asymptotic expansion such as the usual Schwinger-DeWitt 
one, and is valid for any $\sigma$ (contrary to $\sigma$ small), and even 
for $m=0$. It is, more over, the so-called cumulant of the asymptotic
series. Furthermore, along the diagonal,
our coefficients $\tau_n$ are given by a very simple recursion relation and 
thus, once more contrary to the old expansion, they are actually rather easy 
to find to any order (only the first 3 or 4 coefficients in the 
Schwinger-DeWitt expansion can be found in general). It is even
possible to find the coefficients for $x\neq x'$, though this was not done in 
this paper.\\
Some improvements over the Schwinger-DeWitt expansion by other authors should 
be mentioned. First, the approach of Avramidi \cite{Avramidi}, which gives an 
analytical formula for the remainder of the
Schwinger-DeWitt expansion, making it valid for all $\sigma$. It is still, 
however, not as quickly
convergent, nor as easy to find the coefficients for as for the expansion
of the present paper. Basically the same holds for the 
non-local partial
summation approach of Barovinsky et al. \cite{Barovinsky}.\\
Our approach is not in principle limited to one
of the two regions $\partial R \ll R$ and $\partial R \gg R$, but the 
$\sigma$-integral
becomes difficult to handle in the general case, as no analytical expression 
for it exists to the best of the present authors' knowledge. This limitation 
can, however, be overcome in concrete cases by resorting to a numerical 
calculation. As a final note, let us stress
that the accuracy of the chosen approach, due to the simple form of the 
recursion relations, can be systematically improved. \\

We have found curvature induced corrections to the Coleman-Weinberg 
potential for
scalar fields and the generalisation thereof to Dirac and Yang-Mills fields by
means of the heat kernel method. From this we learned that:\\

(1) Even when $m=0$ curvature still induces a mass to the fields,
so that $m_{\rm ren}^2 \neq 0$. 
In the case of the Schwarzschild black hole, we found the renormalised mass 
and coupling
constant of a scalar field only to differ from the flat space values in the 
vicinity of the horizon. Precisely
at the horizon $r=2M$, the renormalised mass was found to be divergent 
expressing the fact that, seen from an external observer (Schwarzschild 
coordinates) an in-falling object never reaches the
horizon. Let us also note
that these curvature induced corrections to the renormalised mass and 
coupling constant that can have either sign {\em a priori}, so curvature can 
both increase and decrease these values. This latter remark also implies that 
under certain circumstances an originally massless field can become tachyonic 
(i.e. $m_{\rm ren}^2 < 0$), perhaps forcing a spontaneous symmetry breaking.\\

(2) Symmetry breaking occurred also in the Yang-Mills case explicitly, 
since we found that the renormalised mass in general was non-zero. Even 
though the classical action is gauge invariant it is well 
known that the effective action need not be \cite{EffAct}. One could take 
this either as a way of spontaneously breaking symmetries and thus producing 
masses without the need of a 
Higgs field or perhaps as an indication of the need, in curved backgrounds, of 
``entangling'' the internal symmetries with the Lorentz transformations, 
i.e. letting the symmetry group not just be $G_{\rm internal}\times 
G_{\rm spacetime}$ but something more complicated. Lending support to the old 
hope that the successful quantisation of gravity
would entail a unification of all forces. It is worth keeping in mind, that 
contributions from the matter fields to the renormalised mass of the gauge 
boson and those from the curvature can cancel each-other under sufficiently 
favourable conditions (symmetry restoration). Spontaneous
 symmetry breaking is of course already present in the mean field
approach by putting $\langle \phi\rangle=\sqrt{\langle \phi^2\rangle}$, a 
breaking which is induced non-perturbatively by the background curvature. \\

(3) A very interesting feature is the appearance in $V_{\rm eff}$ of terms 
which are due to the 
curvature but survive in the limit $\xi R+{\cal E},\xi R+ 2{\cal E}
\rightarrow 0$. These 
terms only appear in $m_{\rm ren}, \lambda_{\rm ren}$, however, provided 
$\partial _p(\xi R+{\cal E}),\partial _p(\xi R+2{\cal E})\not\rightarrow 0$ 
in this limit. For the Yang-Mills field, however, we found that the 
renormalised coupling contains a
manifestly curvature independent non-vanishing term.\\

(4) In the Schwarzschild example the effective action acquires an imaginary 
part, which therefore implies curvature induced particle production.\\

(5) The presence of phase transitions (thought to be responsible for
inflation in some scenarios) will of course be modified by these new terms 
in the effective potential.
In fact we found that a classical critical point was almost never a 
critical point of the effective potential. For a classical critical point 
to ``survive'' the quantisation, both its precise value and the background 
geometry itself had to satisfy certain
extra requirements. For scalar fields we were able to solve these, while we 
for Yang-Mills
fields furthermore found restrictions on the structure of the Lie algebra. We 
also found the critical points of the effective action (i.e., quantum 
critical points) for scalar fields, and made some comments on 
fields of higher spin. It was for instance shown that a Yang-Mills theory 
with $su_2$ as its
gauge algebra could not have a critical point in a general background.\\

(6) The resulting renormalised energy-momentum tensor will in general 
violate the 
null energy condition (NEC) for scalar fields, and have non-vanishing trace 
even for $m=0,\xi=\frac{1}{6}$ (conformal anomaly). For Yang-Mills fields NEC
was {\em not} violated but we still had a conformal anomaly.

\newpage
\begin{table}[htb]
\centering
\begin{tabular}{|l|l||l|}\hline
name & abbrev. & condition \\ \hline
null & NEC & $T_{\mu\nu}k^\mu k^\nu\geq 0~\forall k^\mu\mbox{ null}$\\
weak & WEC &  $T_{\mu\nu}k^\mu k^\nu\geq 0~\forall k^\mu\mbox{ timelike}$\\
strong & SEC & $(T_{\mu\nu}-\frac{1}{2}g_{\mu\nu}T)k^\mu k^\nu \geq 0 ~\forall
 		k^\mu\mbox{ timelike}$\\
dominant & DEC & WEC plus $T_{\mu\nu}k^\mu$ not spacelike\\ \hline
averaged null & ANEC & $\int_\gamma T_{\mu\nu}k^\mu k^\nu ds \geq 0,~\forall k^\mu$ null\\
averaged weak & AWEC & $\int_\gamma T_{\mu\nu}k^\mu k^\nu ds \geq 0,~\forall k^\mu$ timelike\\
averaged strong & ASEC & $\int_\gamma(T_{\mu\nu}k^\mu k^\nu+\frac{1}{2}T)ds \geq 0, ~\forall k^\mu$ timelike\\ \hline
\end{tabular}
\caption{The various energy conditions. For the averaged conditions, $\gamma$ denotes an 
arbitrary curve with (normalized) tangent $k^\mu$ and $s$ the affine parameter (in the 
null case) or the proper time (in the timelike case).}
\end{table}
\newpage

\begin{figure}
\caption{A plot of the real part of $m^2_{\rm ren}$ for a scalar field in a
Schwarzschild background.}
\end{figure}

\begin{figure}
\caption{A plot of the real part of $m^2_{\rm ren}$ for a Yang Mills field in a
Schwarzschild background (the trace of the Cartan-Killing metric has been
put equal to unity).}
\end{figure}

\begin{figure}
\caption{A plot of $V_{\rm eff}''$ as a function of $\lambda,\cal E$ for a
scalar field with a $\phi^4$ self interaction. The positive region is where
the consistency condition is satisfied. The region turn out to be the same
for either choice of sign in the definition of $V_{\rm eff}''$.}
\end{figure}

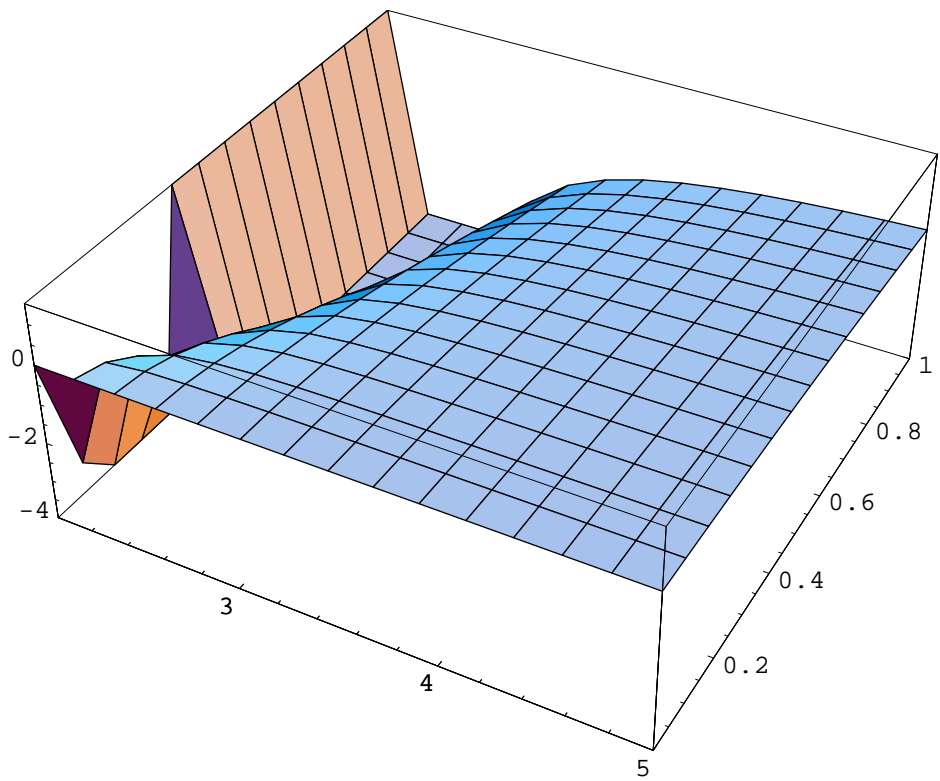
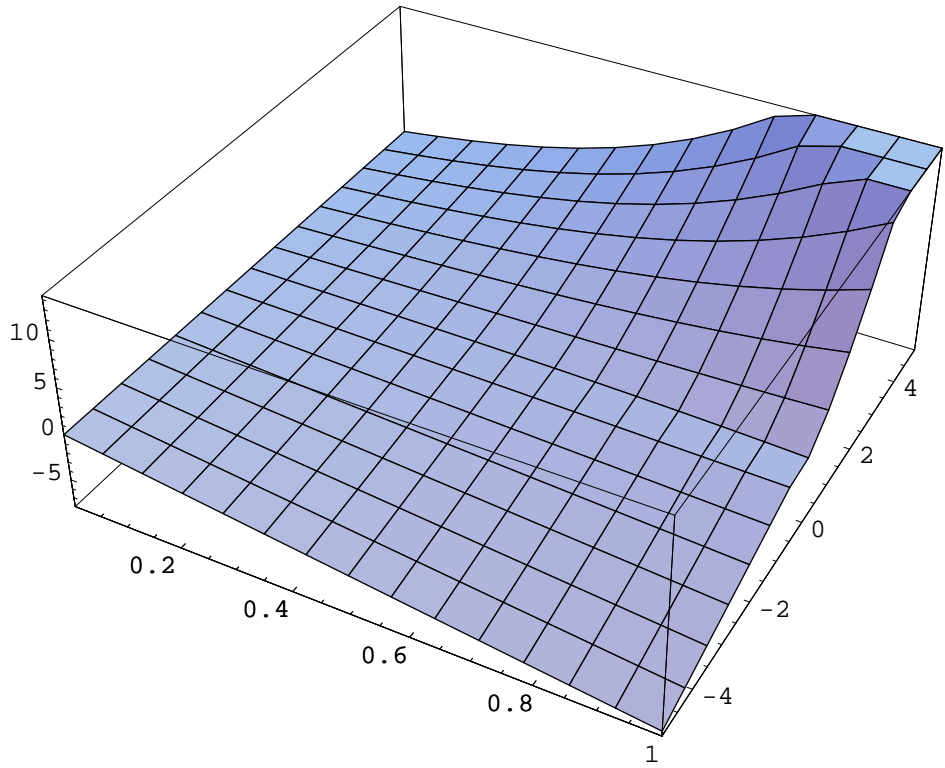
\end{document}